\documentclass[aps,twocolumn,showpacs,preprintnumbers,prl,superscriptaddress,10pt]{revtex4-1}

\usepackage[utf8]{inputenc}
\usepackage{graphicx,amssymb,amsmath,amsthm,amsfonts,epsfig,times,natbib}

\usepackage[usenames,dvipsnames]{color}
\usepackage{epstopdf}
\usepackage{amsmath,amssymb}
\usepackage{tensor}
\usepackage{mathtools}
\usepackage{amsbsy}
\usepackage{bm,url}
\usepackage[linktocpage]{hyperref}
\usepackage{cancel}

\usepackage[scaled]{beramono}
\usepackage[T1]{fontenc}

\begin{document}


\definecolor{orange}{rgb}{0.9,0.45,0}

\newcommand{\re}{\mbox{Re}}
\newcommand{\im}{\mbox{Im}}

\newcommand{\tf}[1]{\textcolor{green}{TF: #1}}
\newcommand{\nsg}[1]{\textcolor{blue}{#1}}
\newcommand{\ch}[1]{\textcolor{red}{CH: #1}}
\newcommand{\fdg}[1]{\textcolor{orange}{FDG: #1}}
\newcommand{\pcd}[1]{\textcolor{magenta}{#1}}
\newcommand{\mz}[1]{\textcolor{cyan}{[\bf MZ: #1]}}

\def\CovDev{D}
\def\Res{{\mathcal R}}
\def\Gammaflat{\hat \Gamma}
\def\metricflat{\hat \gamma}
\def\Dflat{\hat {\mathcal D}}
\def\part_n{\partial_\perp}

\def\Lie{\mathcal{L}}
\def\A{\mathcal{X}}
\def\Aphi{\A_{\phi}}
\def\hAphi{\hat{\A}_{\phi}}
\def\E{\mathcal{E}}
\def\Ham{\mathcal{H}}
\def\M{\mathcal{M}}
\def\R{\mathcal{R}}
\def\p{\partial}

\def\hg{\hat{\gamma}}
\def\hA{\hat{A}}
\def\hD{\hat{D}}
\def\hE{\hat{E}}
\def\hR{\hat{R}}
\def\hcA{\hat{\mathcal{A}}}
\def\hDelt{\hat{\triangle}}

\renewcommand{\t}{\times}

\long\def\symbolfootnote[#1]#2{\begingroup%
\def\thefootnote{\fnsymbol{footnote}}\footnote[#1]{#2}\endgroup}


\title{
Synchronised gravitational atoms  from mergers of bosonic stars
}

\author{Nicolas Sanchis-Gual}
\affiliation{Centro de Astrof\'\i sica e Gravita\c c\~ao - CENTRA, Departamento de F\'\i sica,
Instituto Superior T\'ecnico - IST, Universidade de Lisboa - UL, Avenida
Rovisco Pais 1, 1049-001, Portugal}

    \author{Miguel Zilh\~ao}
\affiliation{Centro de Astrof\'\i sica e Gravita\c c\~ao - CENTRA, Departamento de F\'\i sica,
Instituto Superior T\'ecnico - IST, Universidade de Lisboa - UL, Avenida
Rovisco Pais 1, 1049-001, Portugal}

    \author{Carlos Herdeiro}
\affiliation{Departamento de Matem\'atica da Universidade de Aveiro and 
Centre for Research and Development in Mathematics and Applications (CIDMA), 
Campus de Santiago, 
3810-183 Aveiro, Portugal}

 \author{Fabrizio Di Giovanni}
\affiliation{Departamento de
  Astronom\'{\i}a y Astrof\'{\i}sica, Universitat de Val\`encia,
  Dr. Moliner 50, 46100, Burjassot (Val\`encia), Spain}

\author{Jos\'e~A. Font}
\affiliation{Departamento de
  Astronom\'{\i}a y Astrof\'{\i}sica, Universitat de Val\`encia,
  Dr. Moliner 50, 46100, Burjassot (Val\`encia), Spain}
\affiliation{Observatori Astron\`omic, Universitat de Val\`encia, C/ Catedr\'atico 
  Jos\'e Beltr\'an 2, 46980, Paterna (Val\`encia), Spain}

  \author{Eugen Radu}
\affiliation{Departamento de Matem\'atica da Universidade de Aveiro and 
Centre for Research and Development in Mathematics and Applications (CIDMA), 
Campus de Santiago, 
3810-183 Aveiro, Portugal}




\begin{abstract}
If ultralight bosonic fields exist in Nature as dark matter, superradiance spins down rotating black holes (BHs), dynamically endowing them with equilibrium bosonic clouds, here dubbed \textit{synchronised gravitational atoms} (SGAs). The self-gravity of these same fields, on the other hand, can lump them into (scalar or vector) horizonless solitons known as \textit{bosonic stars} (BSs).  We show that the dynamics of BSs yields a new channel forming SGAs. We study BS binaries that merge to form spinning BHs. After horizon formation, the BH spins up by accreting the bosonic field, but a remnant lingers around the horizon. If just enough angular momentum is present, the BH spin up stalls precisely as the remnant becomes a SGA. Different initial data lead to SGAs with different quantum numbers. Thus, SGAs may  form both from superradiance-driven BH spin \textit{down} and accretion-driven BH spin \textit{up}. The latter process, moreover, can result in heavier SGAs than those obtained from the former: in one example herein, $\sim 18\%$ of the final system's energy and $\sim 50\%$ of its angular momentum remain in the SGA. We suggest that even higher values may occur in systems wherein \textit{both} accretion \textit{and} superradiance  contribute to the SGA formation.
\end{abstract}
 

\maketitle

{\bf {\em Introduction.}}
Dynamical synchronisation occurs in many physical and biological systems. Communities of fireflies or crickets, sets of metronomes or  pendulums, are amongst the examples wherein individual cycles converge to the same phase, if appropriate interactions are present, see $e.g.$~\cite{Mirollo,2002AmJPh..70..992P,strogaatz}.

In Newtonian gravity, dynamical synchronisation occurs in close binary systems~\cite{1981A&A....99..126H}. Tidal interactions tend to synchronise orbital and rotational periods, locking them. In the Earth-Moon system, the latter has reached this equilibrium stage, whereas the Earth is spinning down to meet the longer orbital period. In the Solar system, full synchronisation has been achieved in the lower mass ratio  Pluto-Charon system~\cite{2014Icar..233..242C}. 

In relativistic gravity, synchronisation has been observed to occur in the interaction between spinning black holes (BHs) and bosonic fields. Via superradiance~\cite{Brito:2015oca},  the BH is \textit{spun down} until it locks with the phase dynamics of the bosonic field~\cite{east2017superradiant1,Herdeiro:2017phl}. This \textit{Letter} presents a new  synchronisation channel, through the dynamics of bosonic star (BS) binaries that form a spinning BH. When just enough angular momentum is present in the binary, the final BH \textit{spins up} by accreting the remnant bosonic field after horizon formation, and the process stops when synchronisation is achieved.

{\bf {\em Self-gravitating ultralight bosonic fields (UBFs).}}
Yet unseen UBFs are plausible dark-matter candidates~\cite{Hui:2016ltb}. For masses in the range $10^{-10}-10^{-20}$ eV, UBFs efficiently trigger superradience of astrophysical spinning BHs~\cite{Arvanitaki:2010sy}. The process transfers a fraction of the BH's mass and angular momentum into a bosonic cloud with a slower (phase) angular velocity than that of the spinning horizon. The process stalls when the horizon angular velocity of the spun down BH synchronises with the cloud's angular velocity~\cite{east2017superradiant1,Herdeiro:2017phl}, creating a \textit{synchronised gravitational atom} (SGA)~\cite{Baumann:2019eav}. For complex UBFs, the BH-SGA system is a stationary ``hairy'' BH within the families found in~\cite{Herdeiro:2014goa,Herdeiro:2016tmi,Santos:2020pmh}. 

UBFs form also horizonless self-gravitating solitons, called BSs~\cite{Schunck:2003kk,Liebling:2012fv,Brito:2015pxa}. They can be labelled by their phase oscillation frequency, $\omega$, and ADM mass, $M$, see $e.g.$~\cite{Herdeiro:2017fhv,Herdeiro:2019mbz}. Some spherical, non-spinning BSs are stable, forming dynamically, for both scalar (${\bf{S}}$) and vector ($\bf{V}$, \textit{a.k.a.} Proca) bosonic fields~\cite{Seidel:1990jh,seidel1994formation,sanchis2017numerical,di2018dynamical}. Such BSs can be evolved in binaries~\cite{palenzuela2008orbital,cardoso2016gravitational,bezares2017final, sanchis2019head}. By constrast, spinning BSs are only dynamically robust in the vector case; the scalar stars are transient and develop instabilities~\cite{Sanchis-Gual:2019ljs}. We will show that the evolution of binaries of stable BSs, both scalar and vector, form  SGAs.

{\bf {\em Setup.}}  
Fully non-linear evolutions of BSs were performed in the same Einstein-Klein-Gordon and Einstein-Proca models as in~\cite{Sanchis-Gual:2019ljs}. We studied binary mergers. Using the initial data described in~\cite{bezares2017final, sanchis2019head} two (${\bf{S}}$ or ${\bf{V}}$) BSs are superimposed, separated along the $x$-axis by a coordinate distance $D$ and boosted in opposite directions along the $y$-axis, with velocity $v_y$. The mergers yield  a spinning BH with a bosonic field remnant  (a.k.a. \textit{cloud}) outside the horizon. The latter stores part of the BSs' system initial mass, $M_i$ and angular momentum $J_i$, and its properties depend on $v_y$. Most simulations were performed for non-spinning (${\bf{S}}$ or ${\bf{V}}$) BSs, but mergers of spinning vector BSs (with parallel spins along the $z$-axis) were also studied. In this case, we have taken $v_y=0$ (head-on collisions); the BSs acquire orbital angular momentum due to frame dragging and a spinning BH forms.

We have used the  codes described in~\cite{Cunha:2017wao,Zilhao:2015tya,sanchis2019head,Sanchis-Gual:2019ljs}, within the \textsc{EinsteinToolkit}
infrastructure~\cite{Loffler:2011ay,Zilhao:2013hia,EinsteinToolkit:2019_10} with 
\textsc{Carpet}~\cite{Schnetter:2003rb} for
mesh-refinement,
\textsc{AHFinderDirect}~\cite{Thornburg:2003sf} for finding
apparent horizons, and \textsc{QuasiLocalMeasures}~\cite{Dreyer:2002mx} for extracting BH mass and angular momentum.
Fields are evolved in time using the codes available in the \textsc{Canuda} library~\cite{Canuda_2020_3565475}.
For details see~\cite{Cunha:2017wao,Canuda_2020_3565475,EinsteinToolkit:web,sanchis2019head,Sanchis-Gual:2019ljs}.

\begin{figure}[h!]
\centering
\includegraphics[height=2.1in]{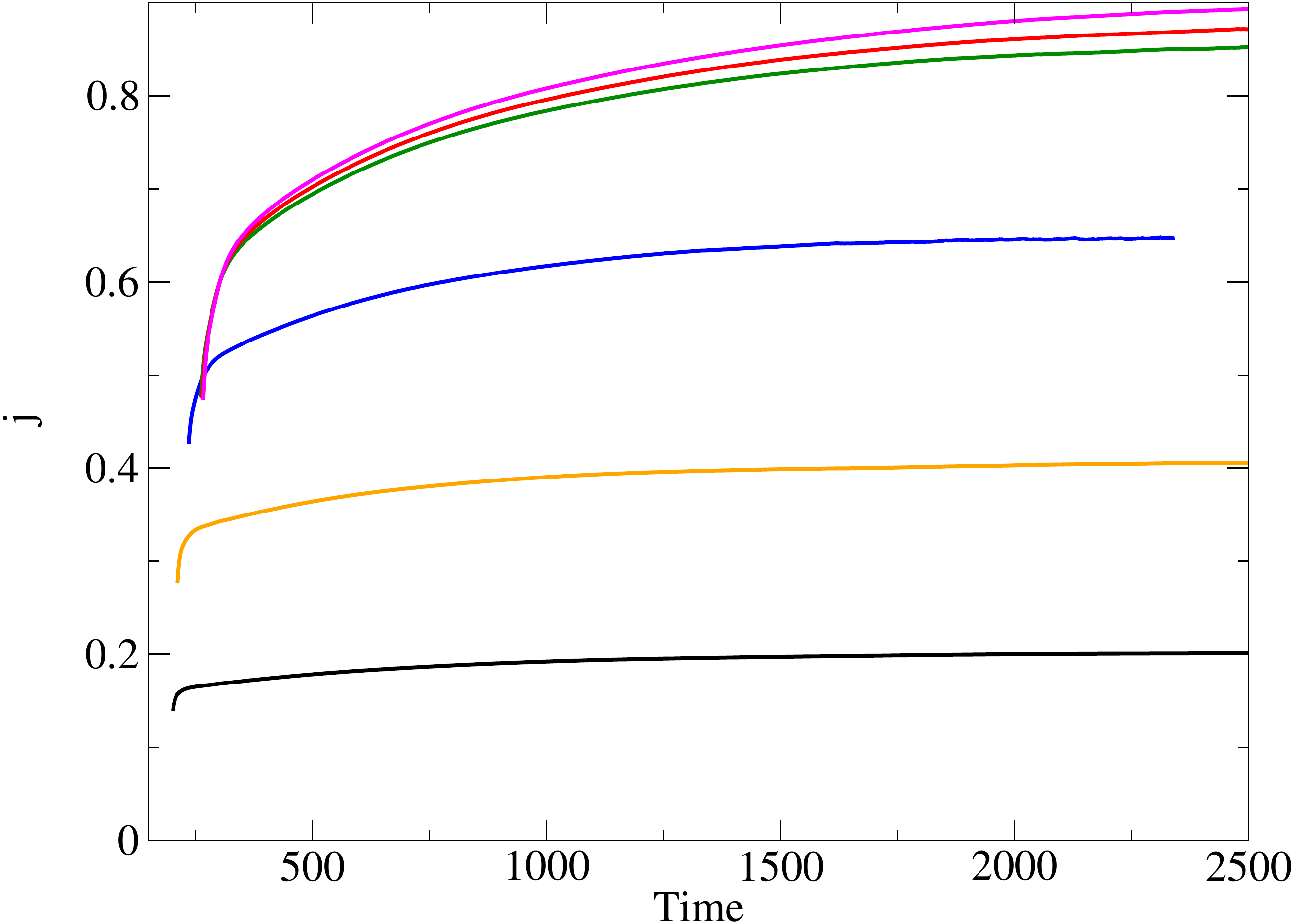}  
\includegraphics[height=2.1in]{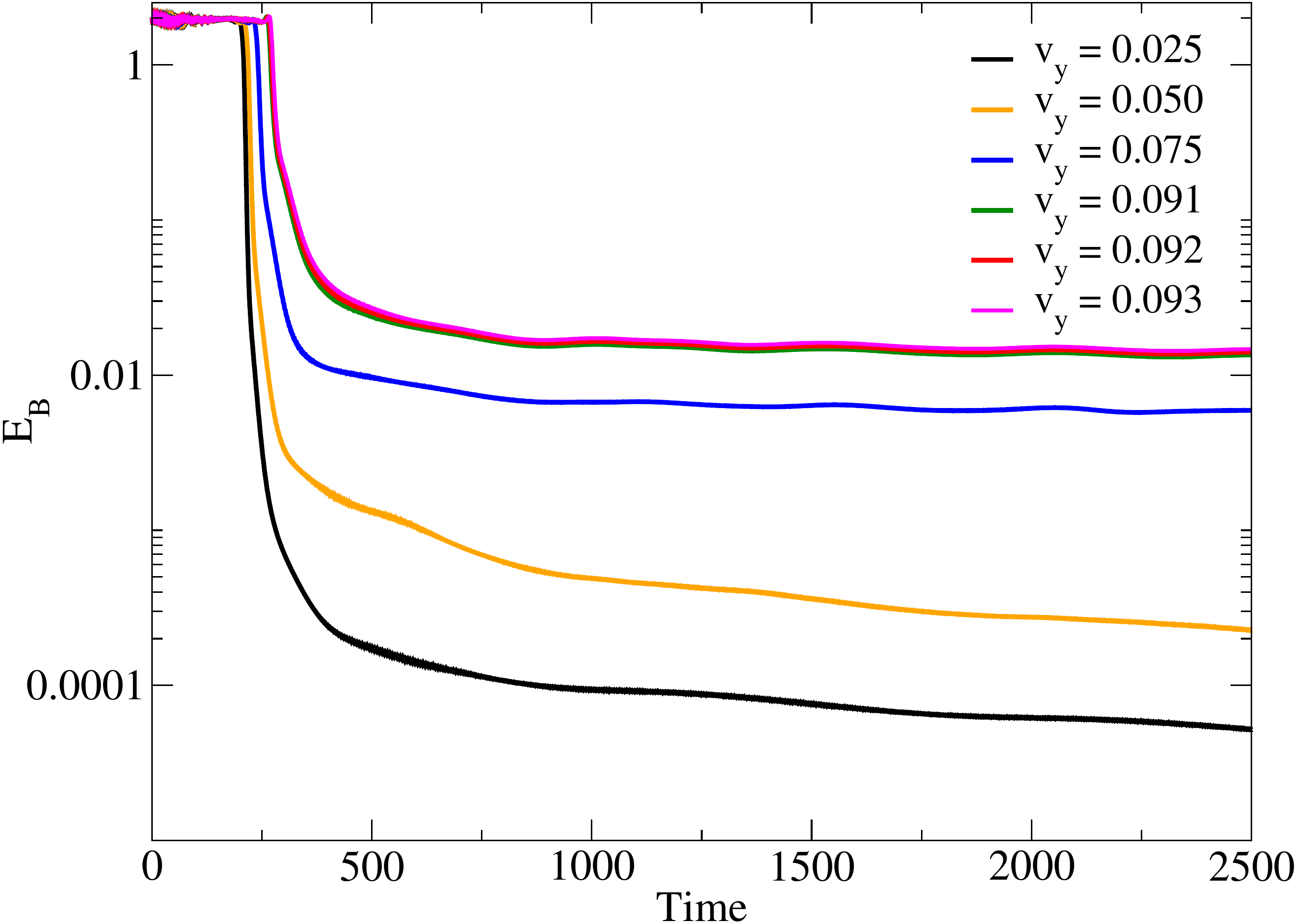}  
\caption{Time evolution of the dimensionless spin of the final BH (top panel) and the energy in the Proca field (bottom panel) for different initial boost velocities, in the merger of two non-spinning vector BSs. 
}
\label{fig1}
\end{figure}

{\bf {\em Mergers of non-spinning BSs.}} 
Two equal-mass binaries of non-spinning BSs were chosen to analyse the remnants after the merger and spinning BH formation: $(i)$ two vector BSs with $(\omega,M)=(0.93,0.952)$~\footnote{Both the Einstein-Klein-Gordon and Einstein-Proca models introduce a single new parameter $\mu$ corresponding to the scalar or vector field mass. This scale is set to unity, so that masses, frequencies, etc, are given in units of $\mu$. Also geometrized units $G= 1 =c$ are used.}; $(ii)$ two scalar BSs with $(\omega,M)=(0.94,0.51)$. The initial separation of the BSs is fixed as $D=30$ (${\bf{V}}$) or $D=16.4$ (${\bf{S}}$). In both cases, the individual BSs are perturbatively stable, and we have performed a number of simulations, varying $v_y$, whose results can be summarised as follows: $1)$ the BSs have an eccentric trajectory, merge and an apparent horizon forms. $2)$ The final BH retains the largest part of $M_i, J_i$, denoted as $M_{\rm BH} ,J_{\rm BH}$, respectively. Thus, the final object is \textit{approximately} a vacuum Kerr BH. Its dimensionless spin $j\equiv J_{\rm BH}/M_{\rm BH}^2$  grows with time after horizon formation due to accretion of the bosonic remnant, saturating at a maximum value; this final $j$ grows with $v_y$ - Fig.~\ref{fig1}, top panel. $3)$ After saturation there is a bosonic cloud outside the horizon, retaining a small fraction of $M_i, J_i$ - Fig.~\ref{fig1}, bottom panel.  Increasing $v_y$, these fractions increase. The energy and spin in the bosonic fields are denoted, respectively $E_{\rm B}, J_{\rm B}$. Initially, $M_i=M_{\rm B}$ and $J_i= J_{\rm B}$. These behaviours are illustrated for the vector case in Fig.~\ref{fig1}. A similar behaviour is found for the scalar case (and for $J_B$, in the case of the bottom panel).
%

\begin{figure}[h!]
\centering
\includegraphics[height=2.1in]{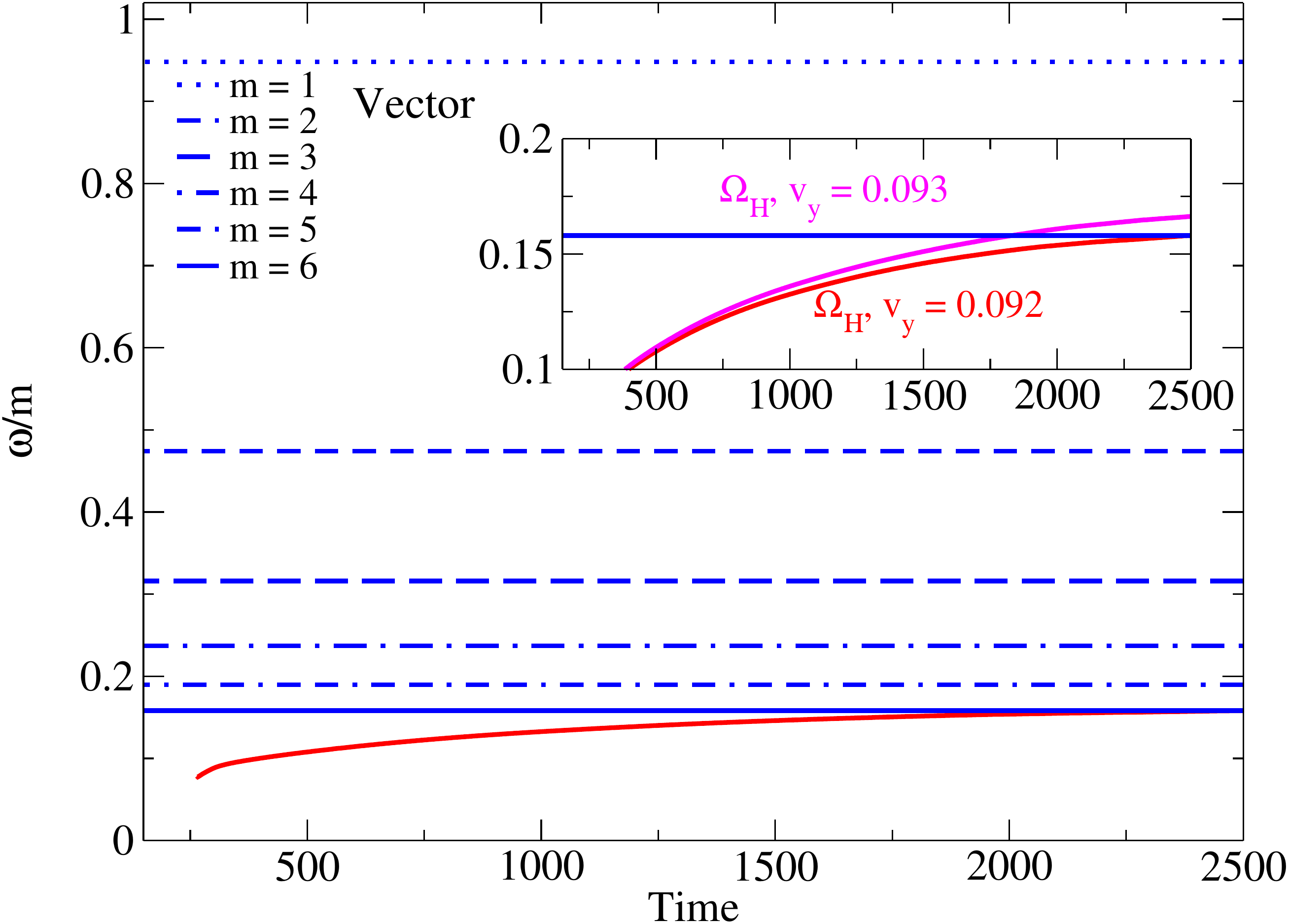} 
\includegraphics[height=2.1in]{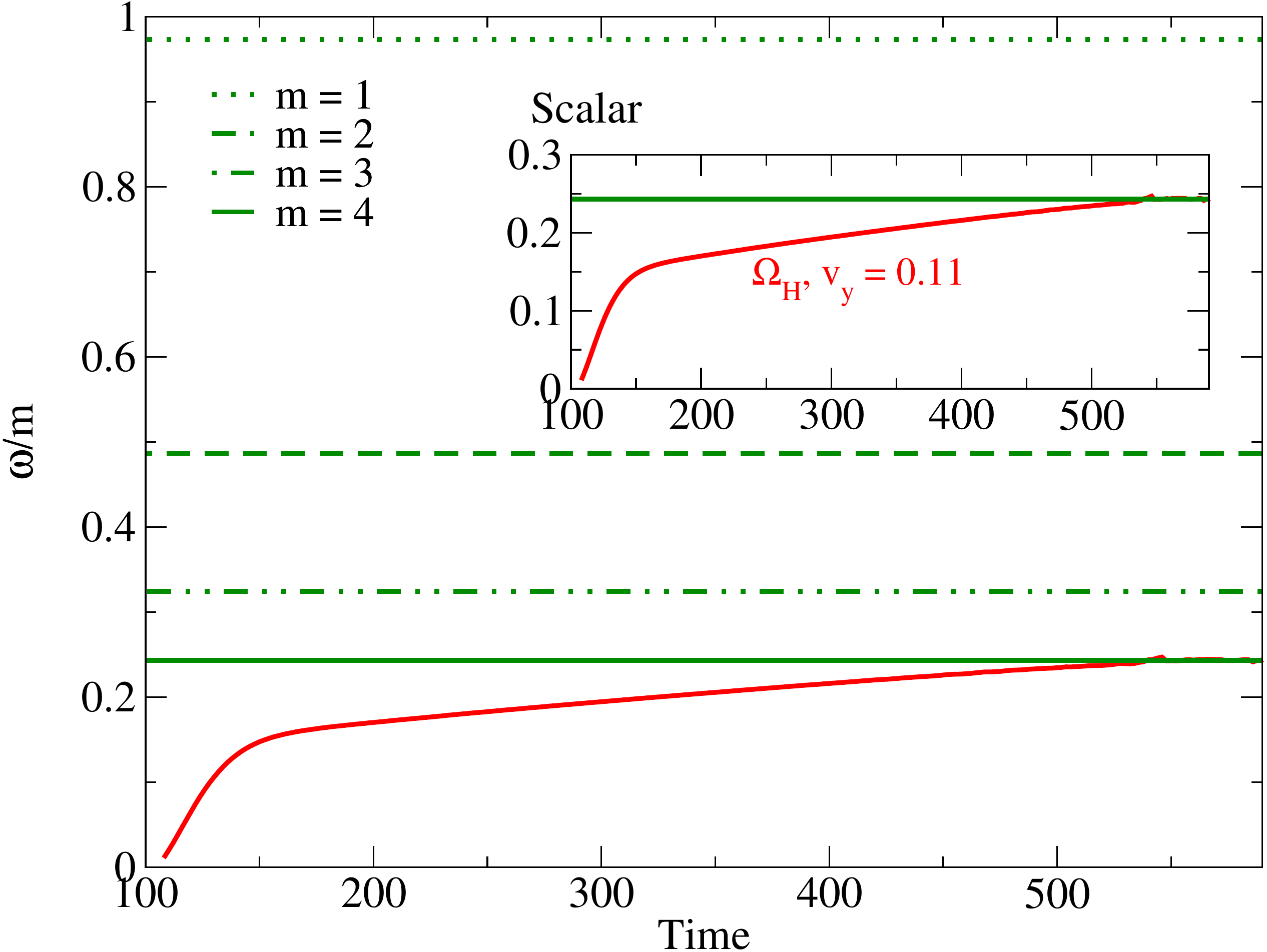}  
\caption{Time evolution of the BH's  $\Omega_H$  in the simulations with $v_y=0.092$ (${\bf{V}}$, top panel) and   $v_y=0.11$ (${\bf{S}}$, bottom panel).  The horizontal lines are the oscillation frequency of the bosonic cloud remnant divided by different $m$'s. 
}
\label{fig2}
\end{figure}

Let us address the nature of the remnant bosonic cloud. Synchronisation in the BH-cloud system occurs when the latter is an oscillating field with phase $\sim e^{-i(\omega t-m \varphi)}$ and the phase angular velocity locks with the BH horizon angular velocity $\Omega_{H}$: $\omega/m=\Omega_H$~\cite{Hod:2012px,Herdeiro:2014goa}. $m\in \mathbb{Z}$ is the azimuthal number of the bosonic remnant. Since the final BH in these simulations is approximately Kerr, we use the standard Kerr relation for $\Omega_H=\Omega_H(M_{\rm BH},J_{\rm BH})$~\cite{Townsend:1997ku}; thus
\begin{equation}\label{vhorizon}
\frac{\omega}{m}=\Omega_{H}\simeq \frac{J_{\rm BH}}{2M_{\rm BH}[M_{\rm BH}^{2}+\sqrt{M_{\rm BH}^{4}-J_{\rm BH}^{2}}]} \ .
\end{equation}
The evolution of $\Omega_H$ is shown in Fig.~\ref{fig2} for the simulations with $v_y=0.11$ (${\bf{S}}$) and $v_y=0.092$ (${\bf{V}}$). 

%
The panels exhibit the spin up of the formed  BH, sourced by the accretion of (part) of the bosonic remnant. They also show the leading oscillation frequency of the remnant bosonic cloud, obtained as a Fourier transform, $\omega=0.973$ (${\bf S}$) or $\omega=0.948$ (${\bf V}$), divided by different values of $m$, corresponding to the different horizontal lines. One concludes that, for these initial data, the spin up stops when the synchronisation condition~\eqref{vhorizon} becomes satisfied, for an $m=6$ (${\bf{V}}$) or $m=4$ (${\bf{S}}$) bosonic cloud remnant (see the insets in Fig.~\ref{fig2}). 
Are these the correct $m$'s that describe the remnants obtained in the simulations?

\begin{figure}[t!]
\includegraphics[width=0.24\linewidth]{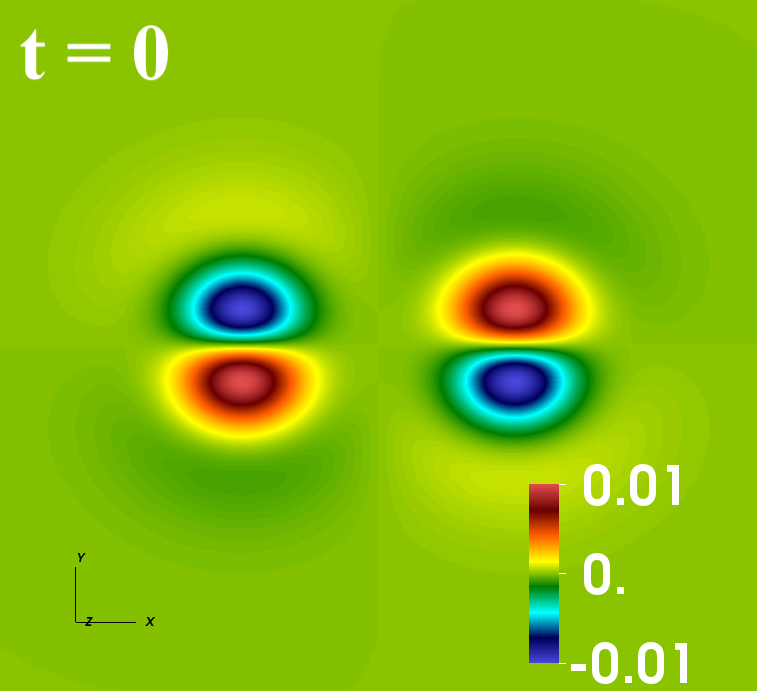}
\includegraphics[width=0.24\linewidth]{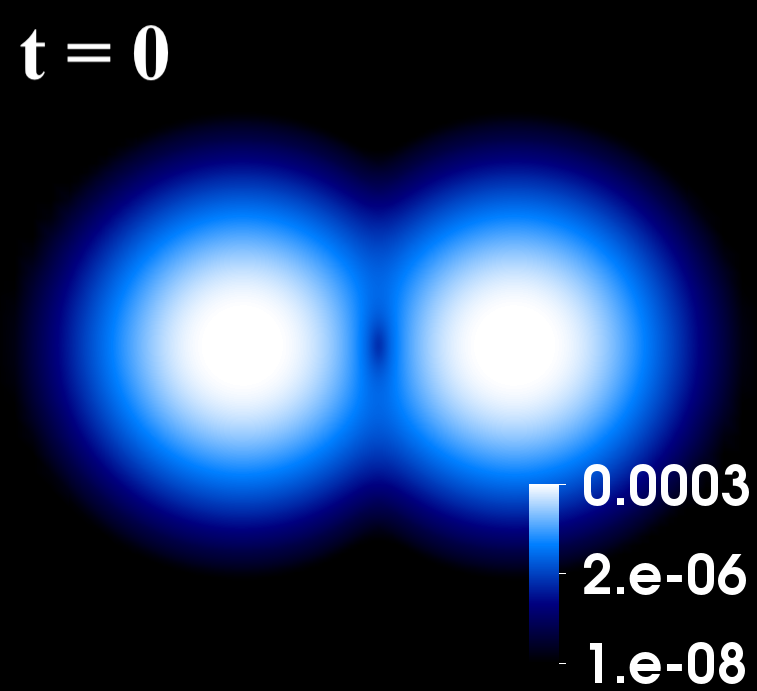}
\includegraphics[width=0.24\linewidth]{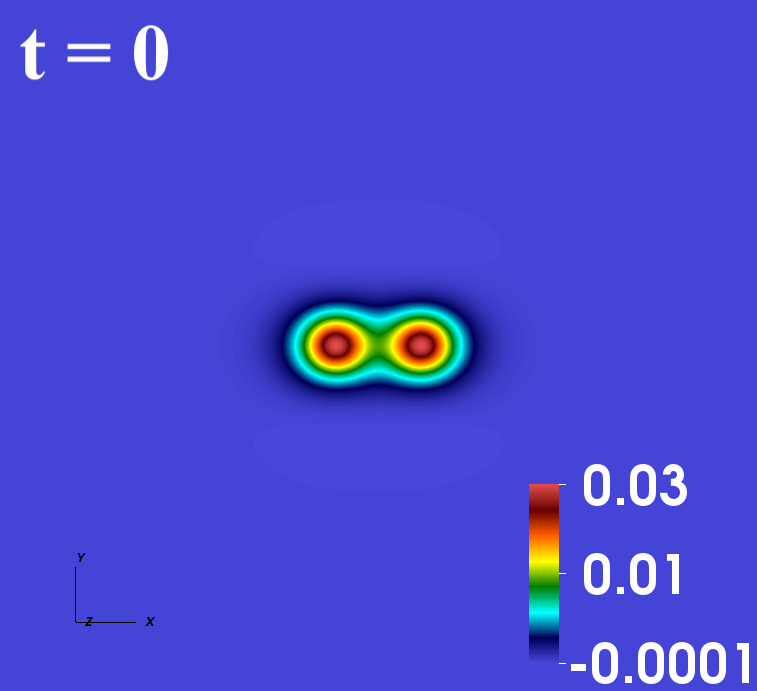}
\includegraphics[width=0.24\linewidth]{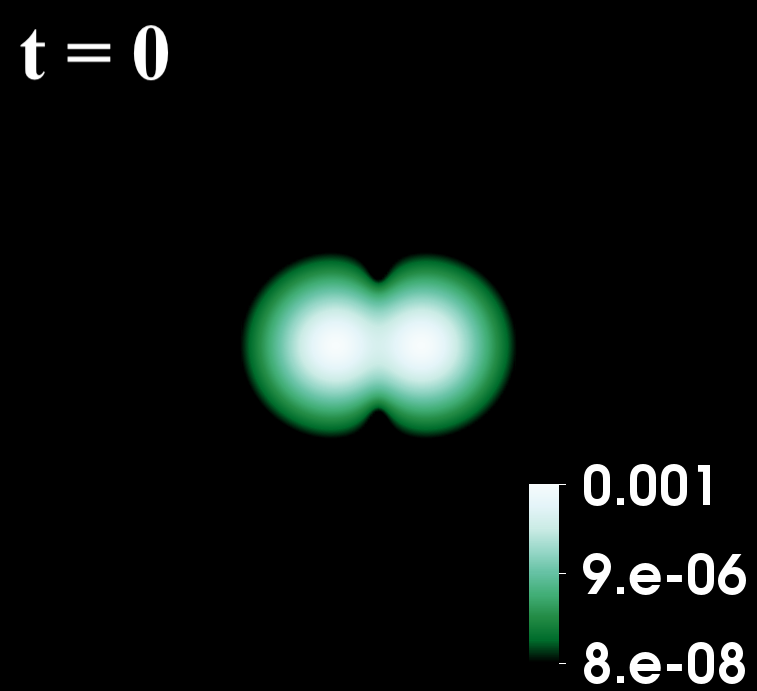}\\
\includegraphics[width=0.24\linewidth]{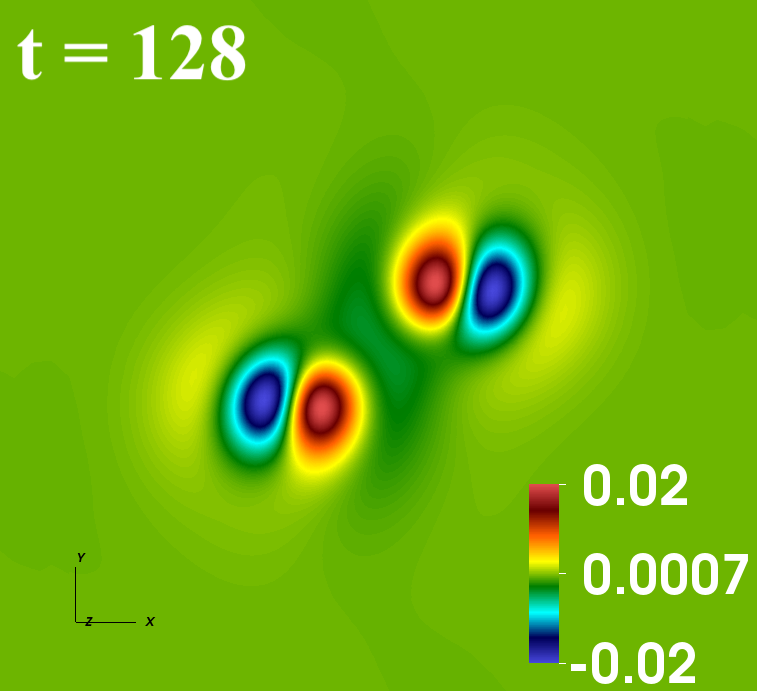}
\includegraphics[width=0.24\linewidth]{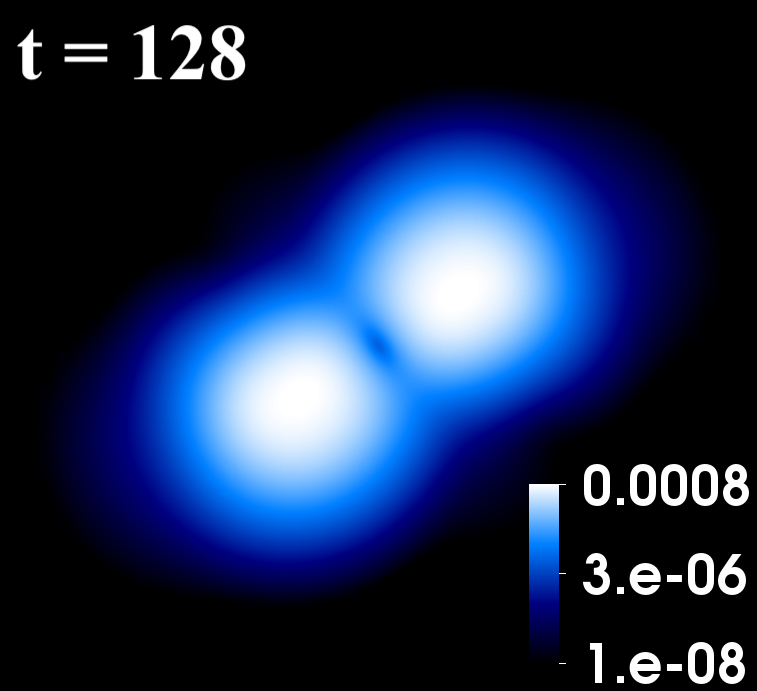}
\includegraphics[width=0.24\linewidth]{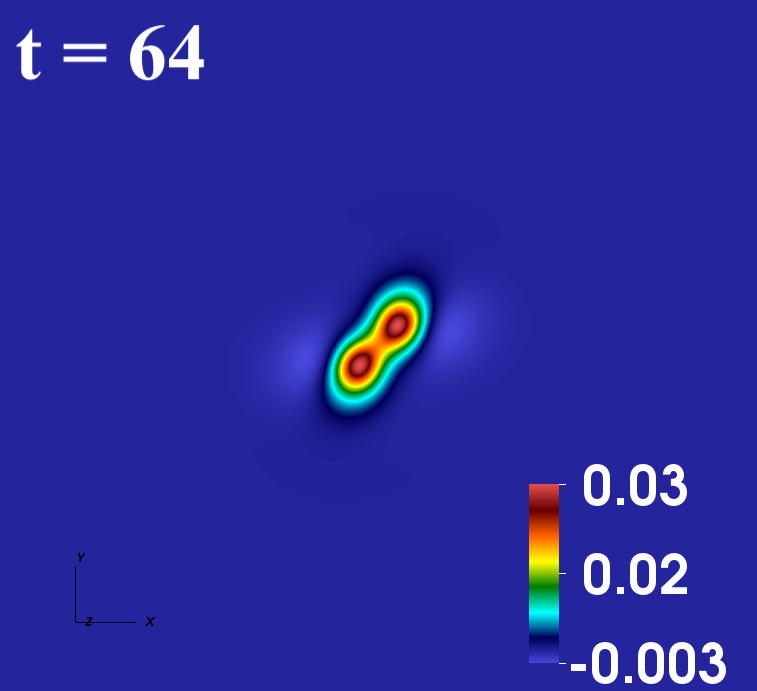}
\includegraphics[width=0.24\linewidth]{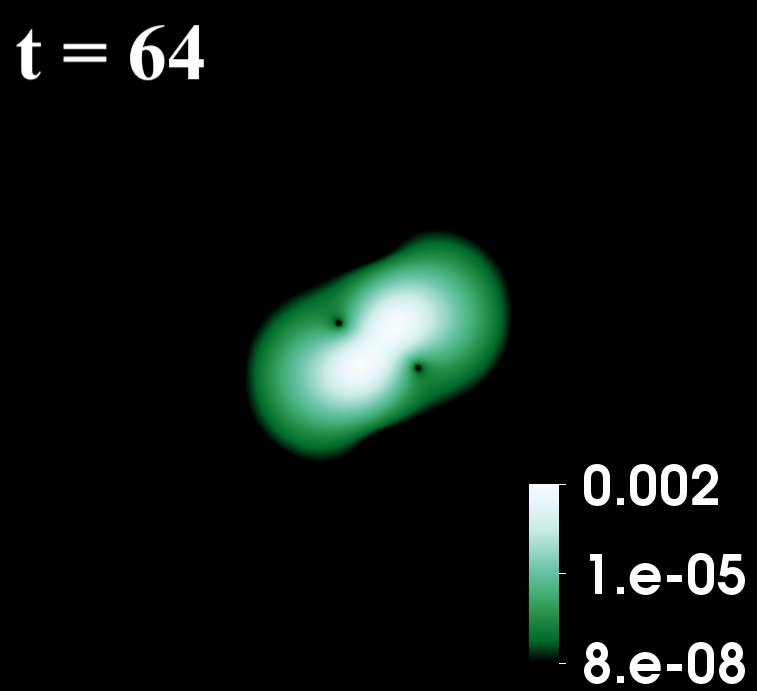}\\
\includegraphics[width=0.24\linewidth]{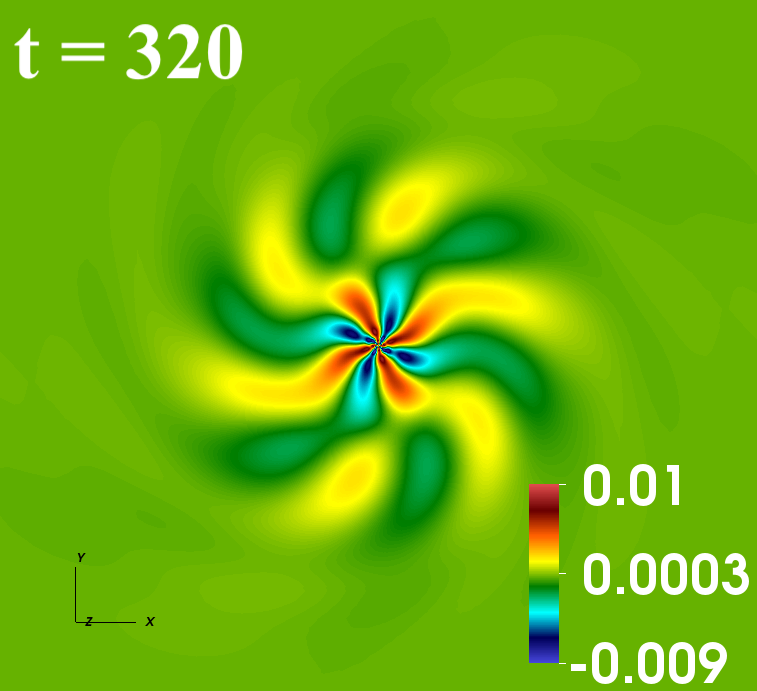}
\includegraphics[width=0.24\linewidth]{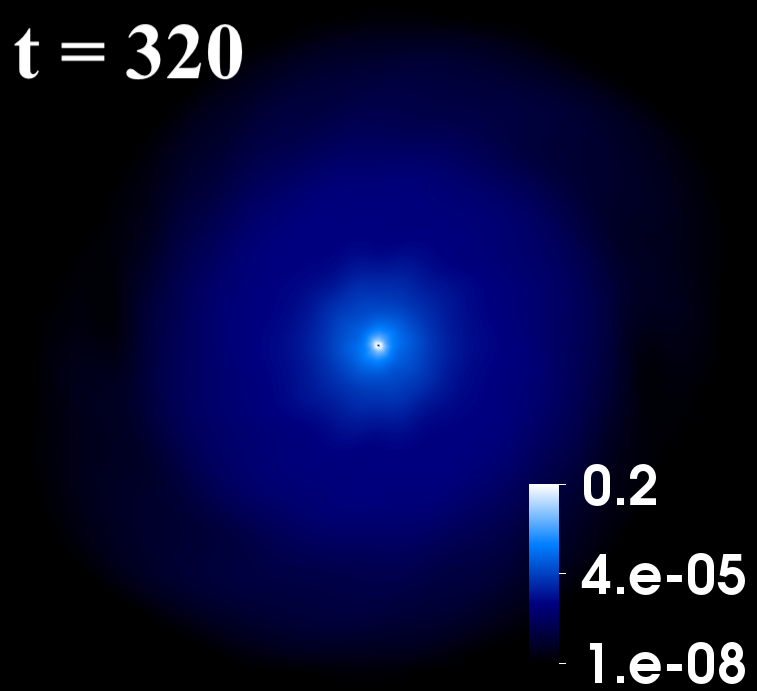}
\includegraphics[width=0.24\linewidth]{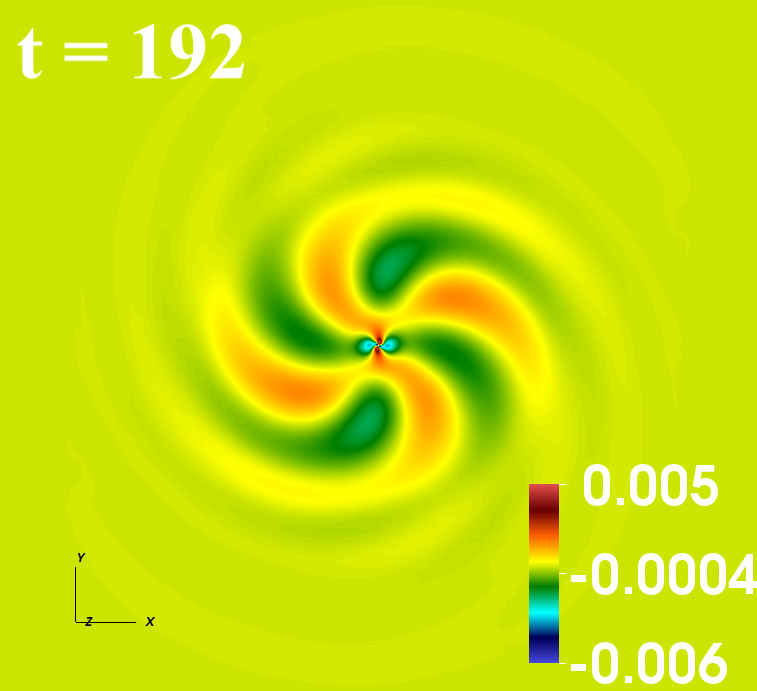}
\includegraphics[width=0.24\linewidth]{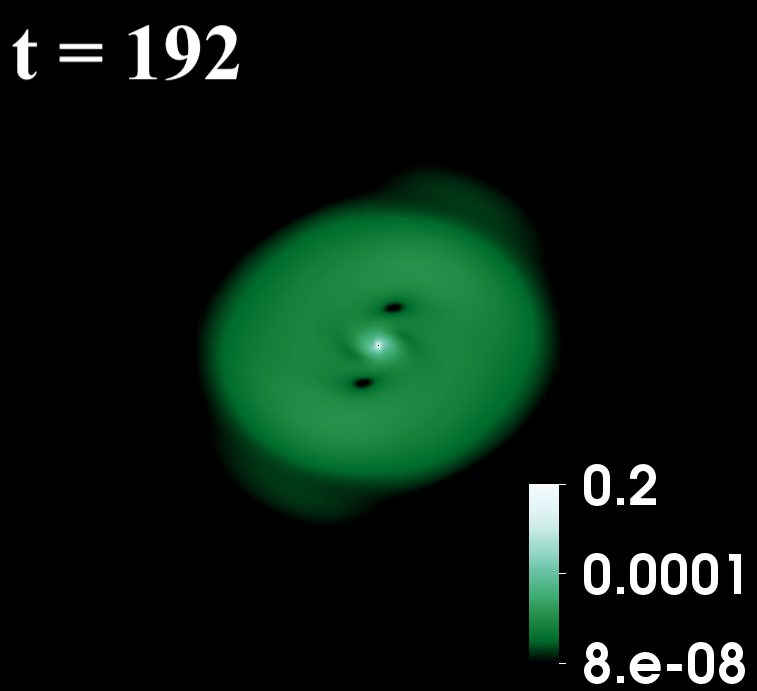}\\
\includegraphics[width=0.24\linewidth]{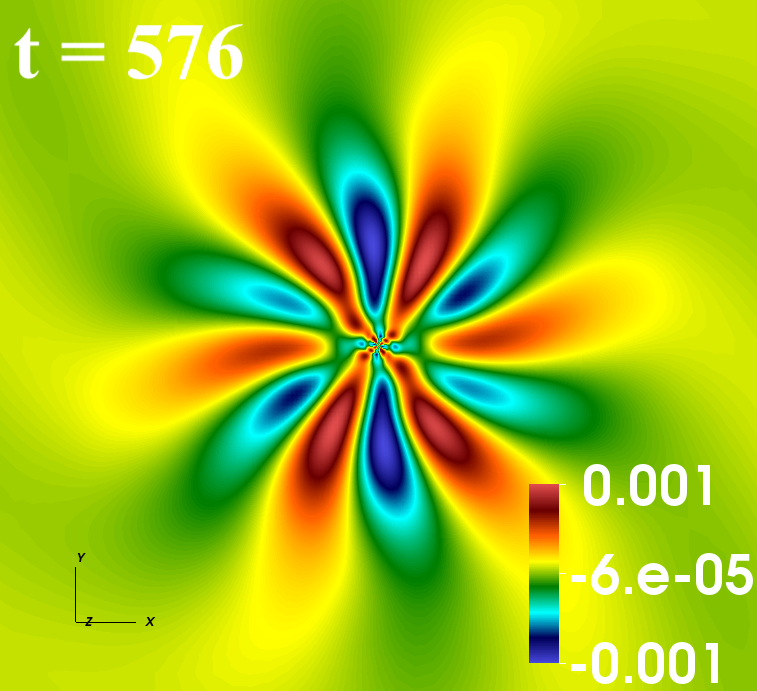}
\includegraphics[width=0.24\linewidth]{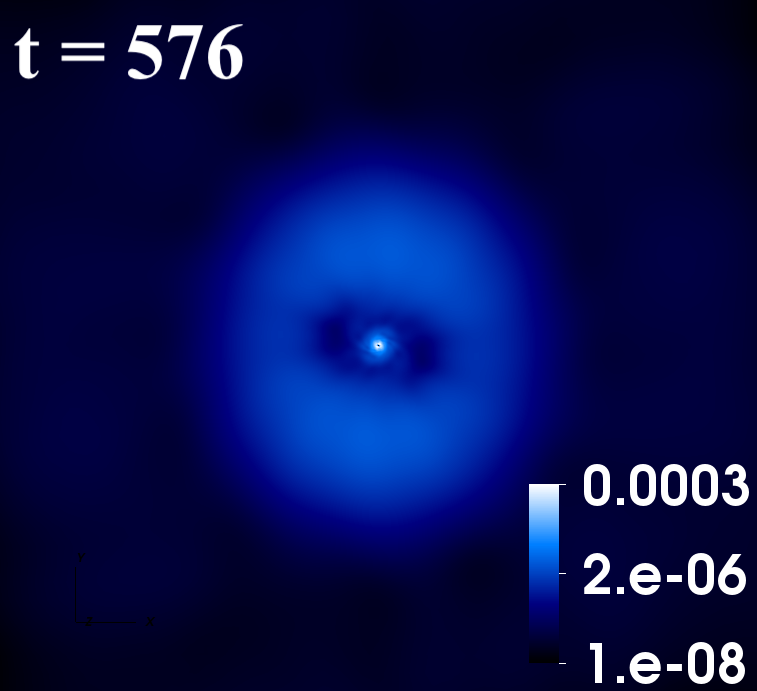}
\includegraphics[width=0.24\linewidth]{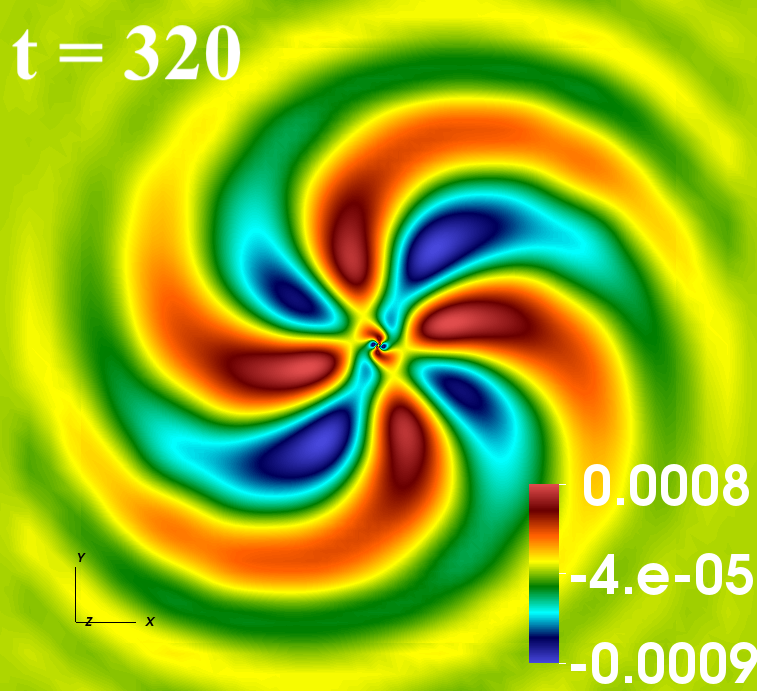}
\includegraphics[width=0.24\linewidth]{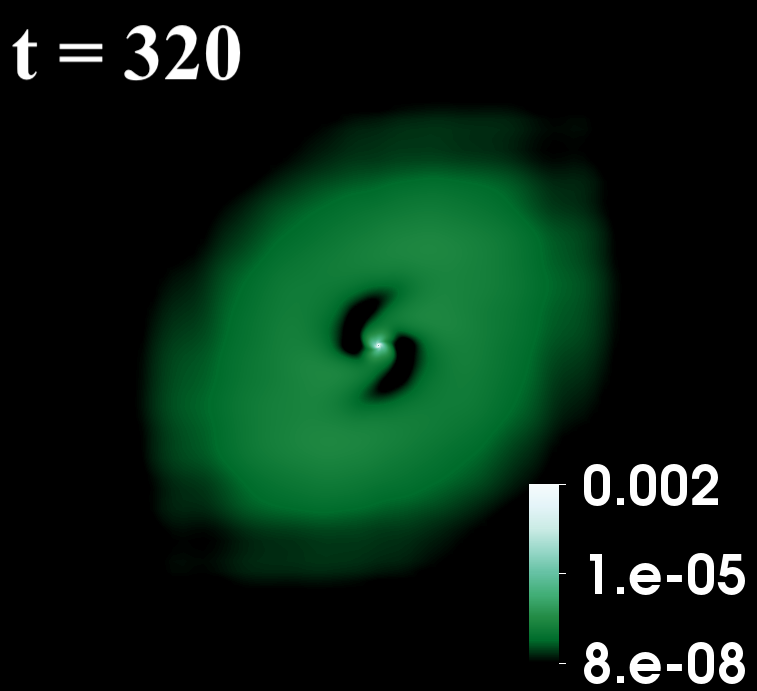}\\
\includegraphics[width=0.24\linewidth]{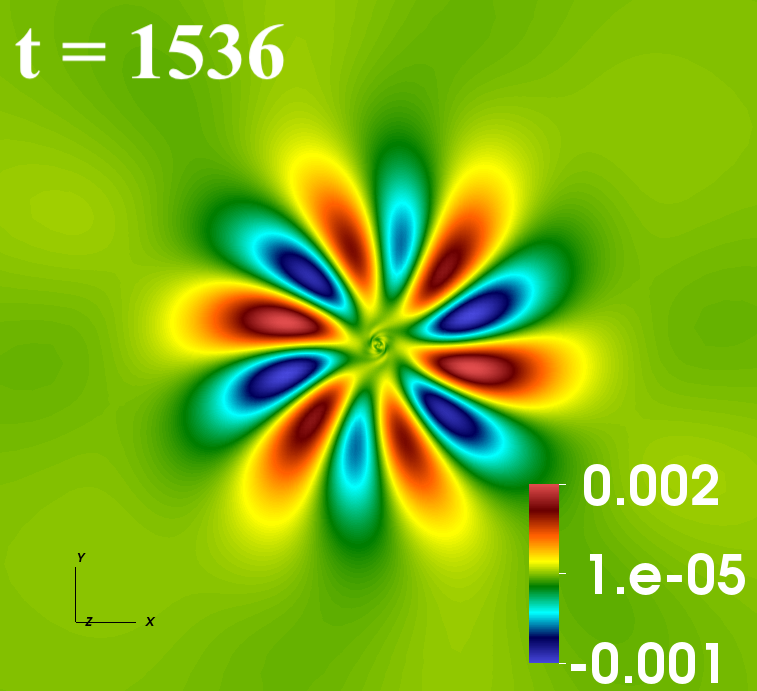}
\includegraphics[width=0.24\linewidth]{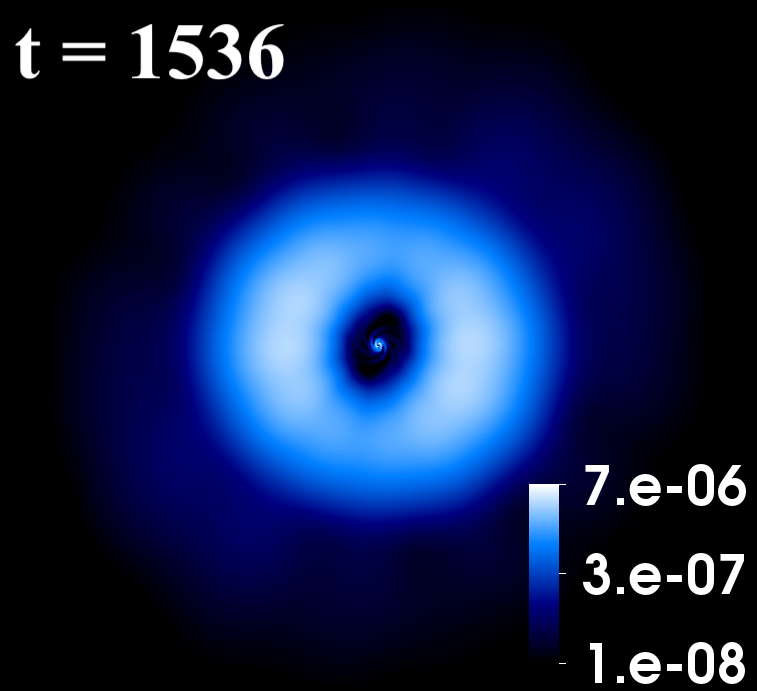}
\includegraphics[width=0.24\linewidth]{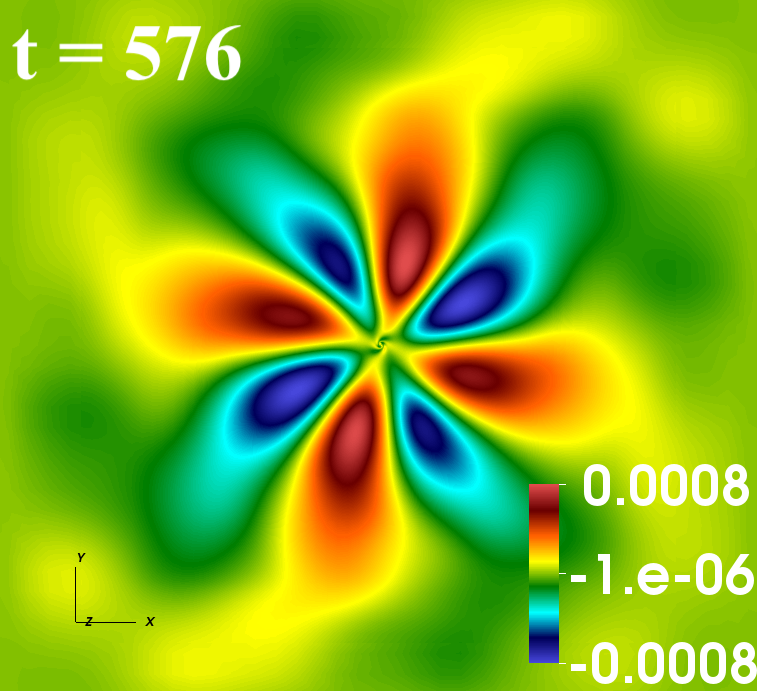}
\includegraphics[width=0.24\linewidth]{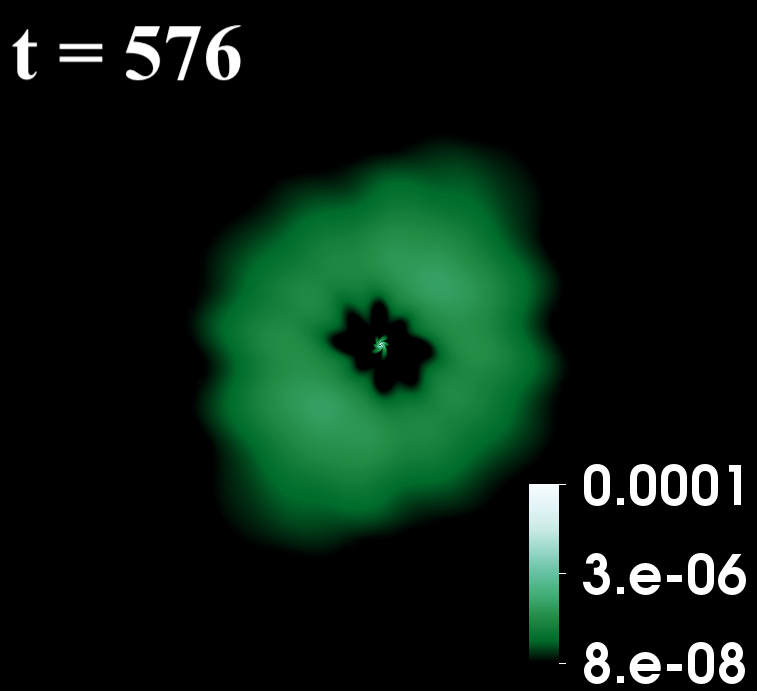}\\
\hspace{-0.01\linewidth}
\includegraphics[width=0.24\linewidth]{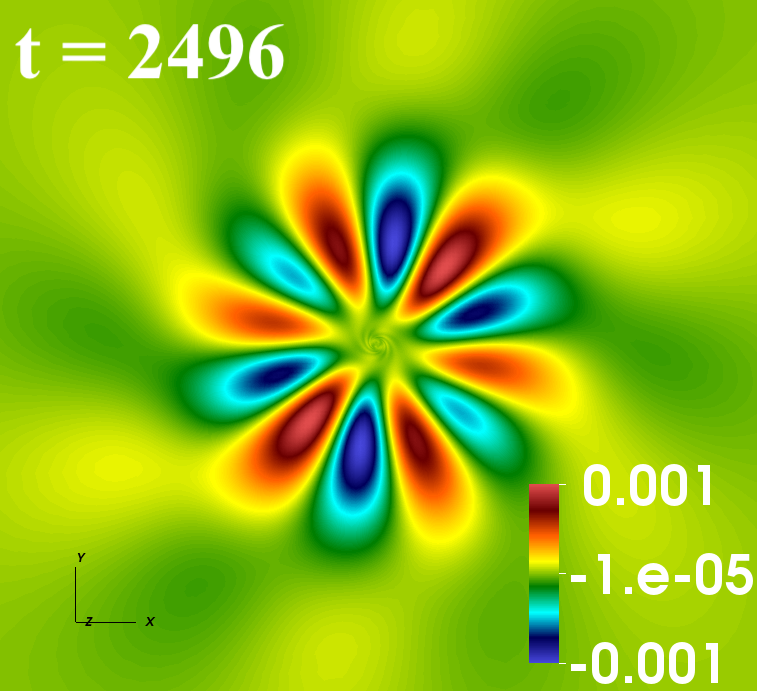}
\includegraphics[width=0.24\linewidth]{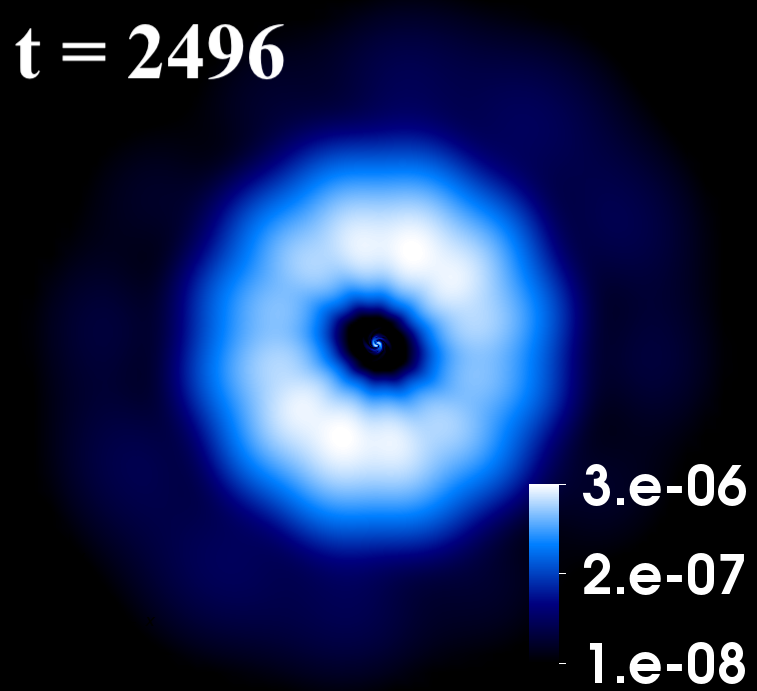}
\includegraphics[width=0.24\linewidth]{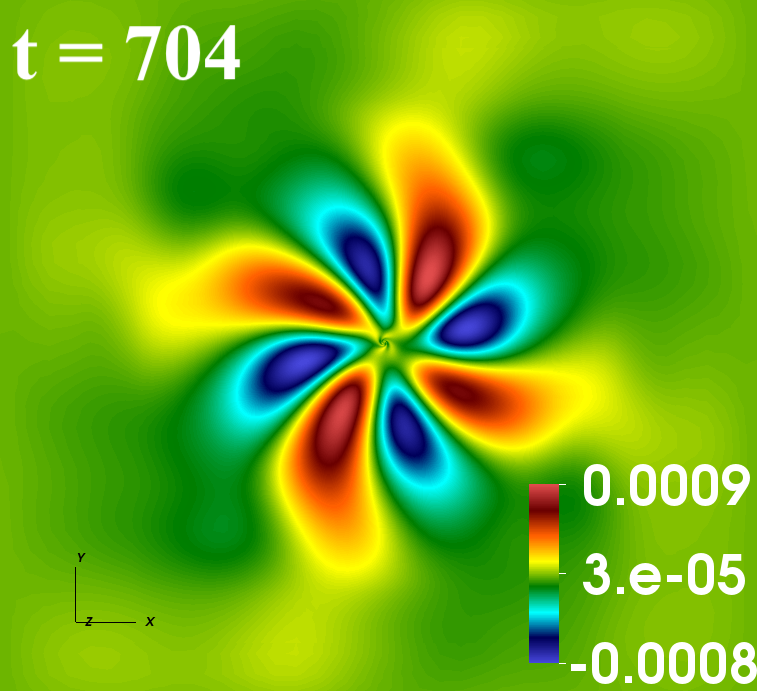}
\includegraphics[width=0.24\linewidth]{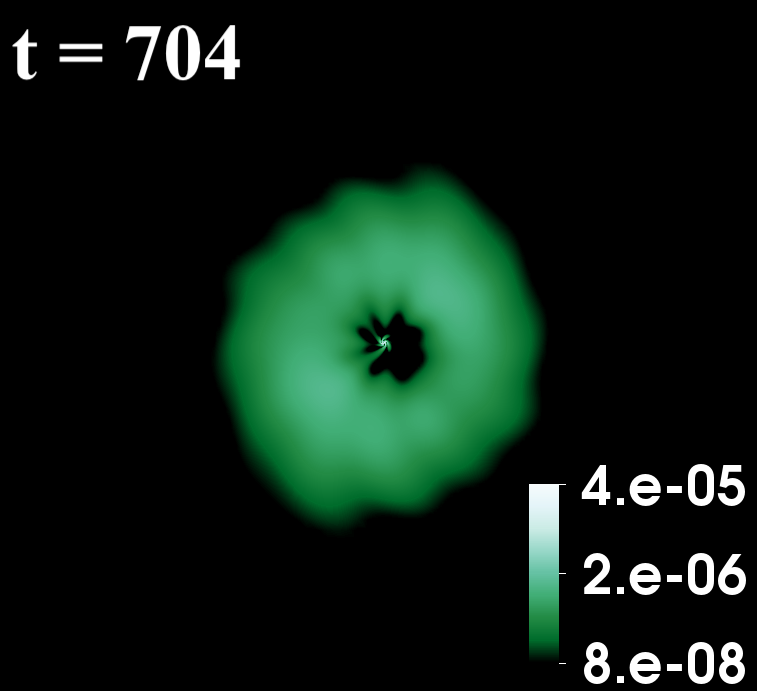}\\
\caption{Equatorial ($xy$) plane snapshots taken during the time evolution of the mergers of non-spinning BSs. (Left column) real part of $\mathcal{X}_{\phi}$; (middle left column) Proca energy density; (middle right column) real part of $\phi$; (right column) scalar energy density.}
\label{fig3}
\end{figure}

\begin{figure}[h!]
\centering
\includegraphics[height=1.53in]{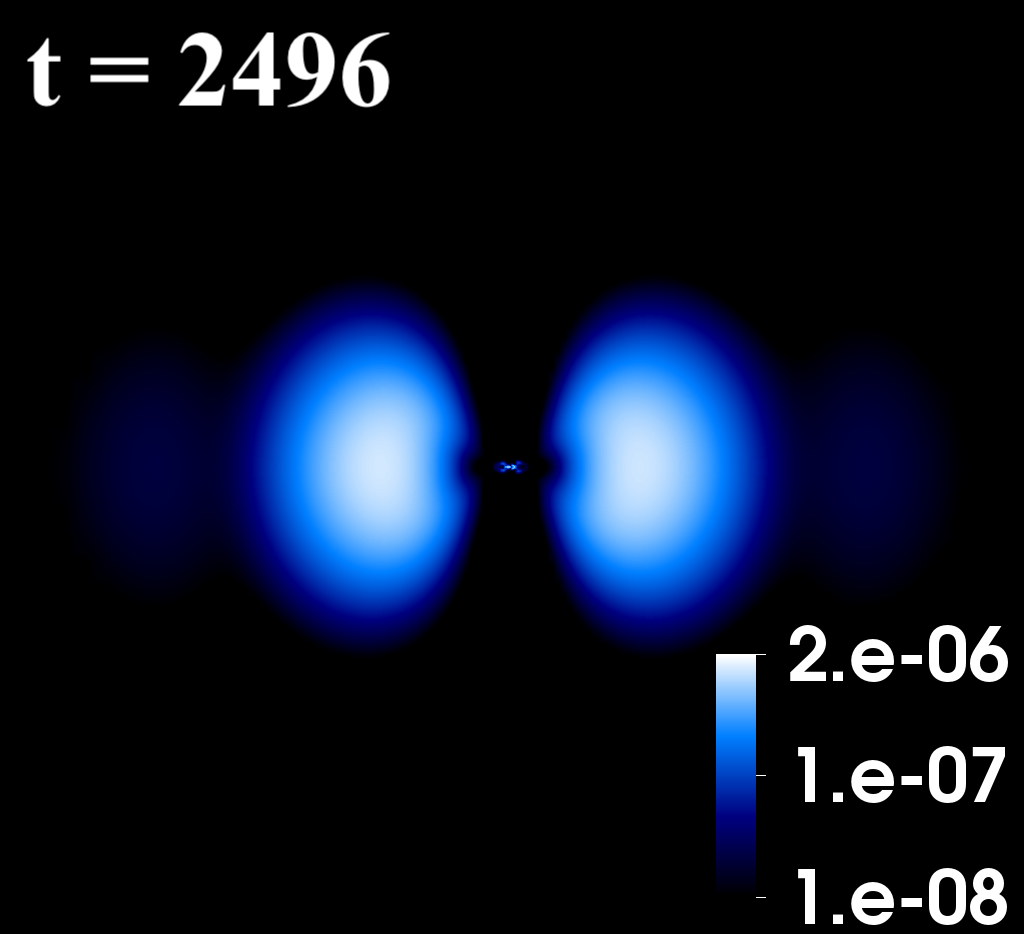}
\includegraphics[height=1.53in]{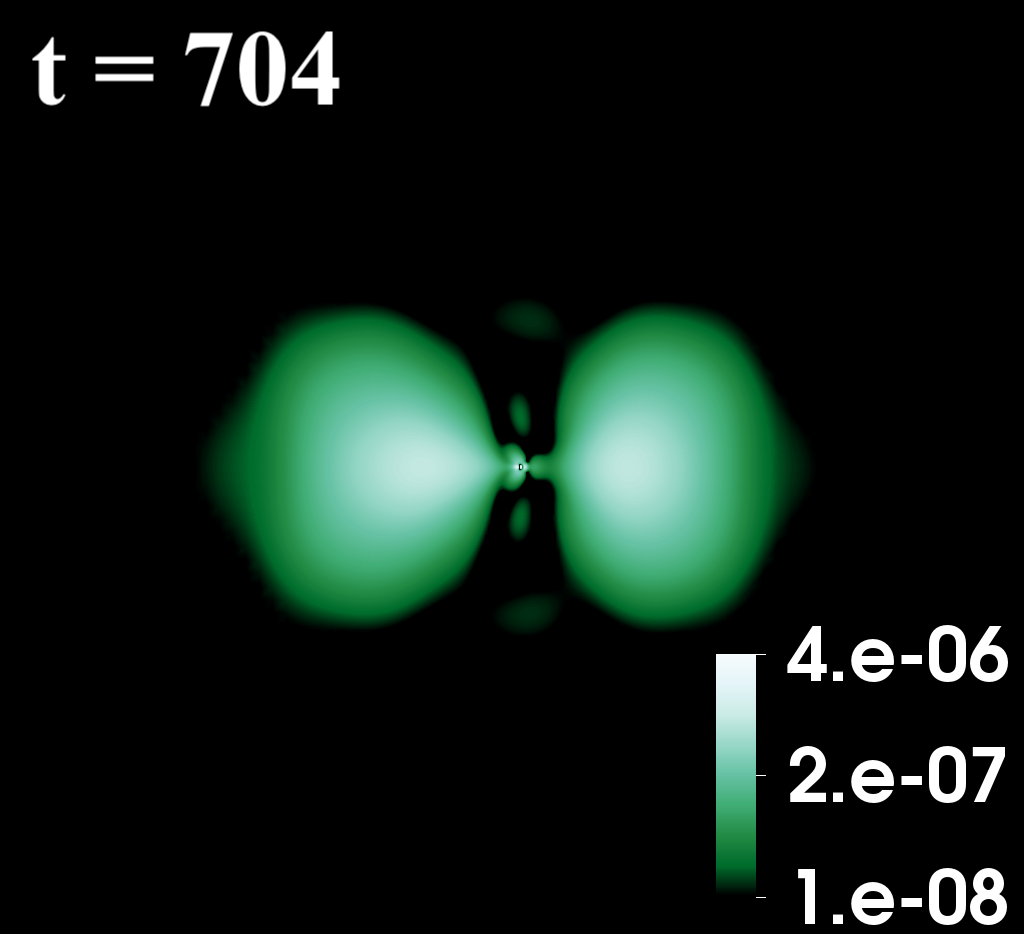}   
\caption{$xz$-plane snapshots of the energy density taken during the time evolution of the mergers of non-spinning BSs for the vector (left panel) and scalar (right panel) cases.}
\label{fig4}
\end{figure}

An affirmative answer is provided in Fig.~\ref{fig3}. Equatorial plane snapshots of the time evolution of both the vector and scalar amplitudes and energy densities are shown. Concerning the amplitudes, the left and middle right columns exhibit the real part of the scalar Proca potential, $\mathcal{X}_{\phi}$~\footnote{See $e.g.$~\cite{sanchis2019head} for the definition of this potential.} and of the scalar field, $\phi$, respectively. The $m=6$ and $m=4$ azimuthal distributions are clearly seen, after the merger. This confirms the bosonic cloud remnant is a synchronised cloud (or SGA) with  $m=6$ (${\bf{V}}$) or $m=4$ (${\bf{S}}$). Concerning the energy densities, one observes in the middle left (${\bf{V}}$) and right (${\bf{S}}$) columns the toroidal shape of the clouds's energy distribution, which is confirmed in Fig.~\ref{fig4}. The inset of Fig.~\ref{fig2} (top panel) also shows a simulation with $v_y=0.093$, whose implications will be discussed below.

\begin{figure}[h!]
\centering
\includegraphics[height=2.1in]{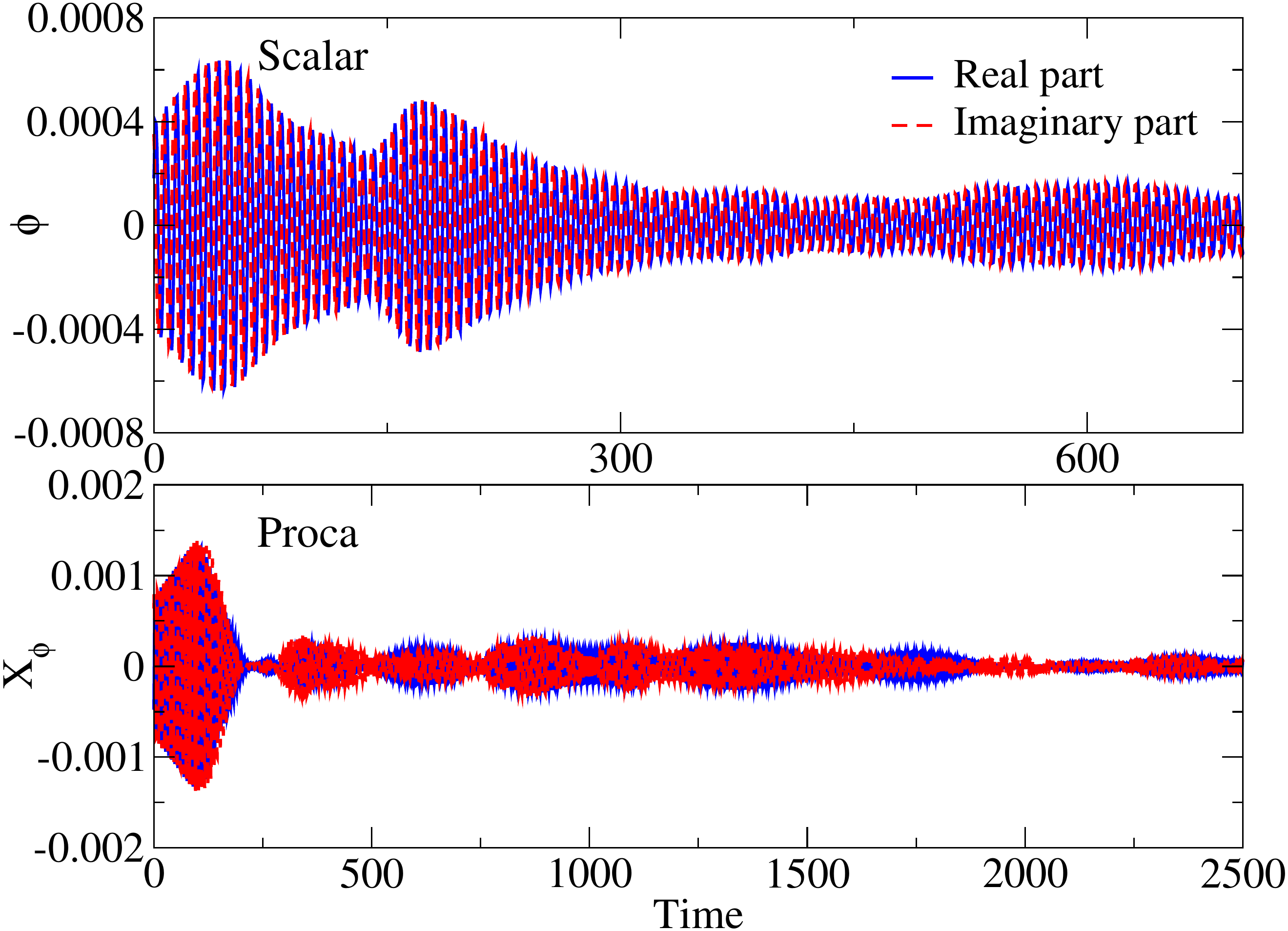}
\caption{Time evolution of the real and imaginary parts of $\phi$ (top panel) and $\mathcal{X}_{\phi}$ (bottom panel), extracted at radius $r=20.78$.
}
\label{fig5}
\end{figure}

If the remnant cloud were a pure synchronised mode it would be stationary.  Only quasi-stationary clouds, however, are obtained from the BS mergers, due to subleading modes.  This is corroborated by Fig.~\ref{fig5}, wherein the time evolution of the amplitude of the real and imaginary parts of $\mathcal{X}_{\phi}$, and $\phi$ are shown for the simulations with $v_y=0.11$ (${\bf{S}}$) and $v_y=0.092$ (${\bf{V}}$). In the vector case, the two leading modes have frequencies, $\omega_{1}=0.948$ (dominant) and $\omega_{2}=0.962$, which produce the beating pattern. In the scalar case, the leading modes have frequencies  $\omega_1 = 0.973$ (dominant) and $\omega_2=0.993$. 
%

\begin{figure}[h!]
\centering
\includegraphics[height=2.3in]{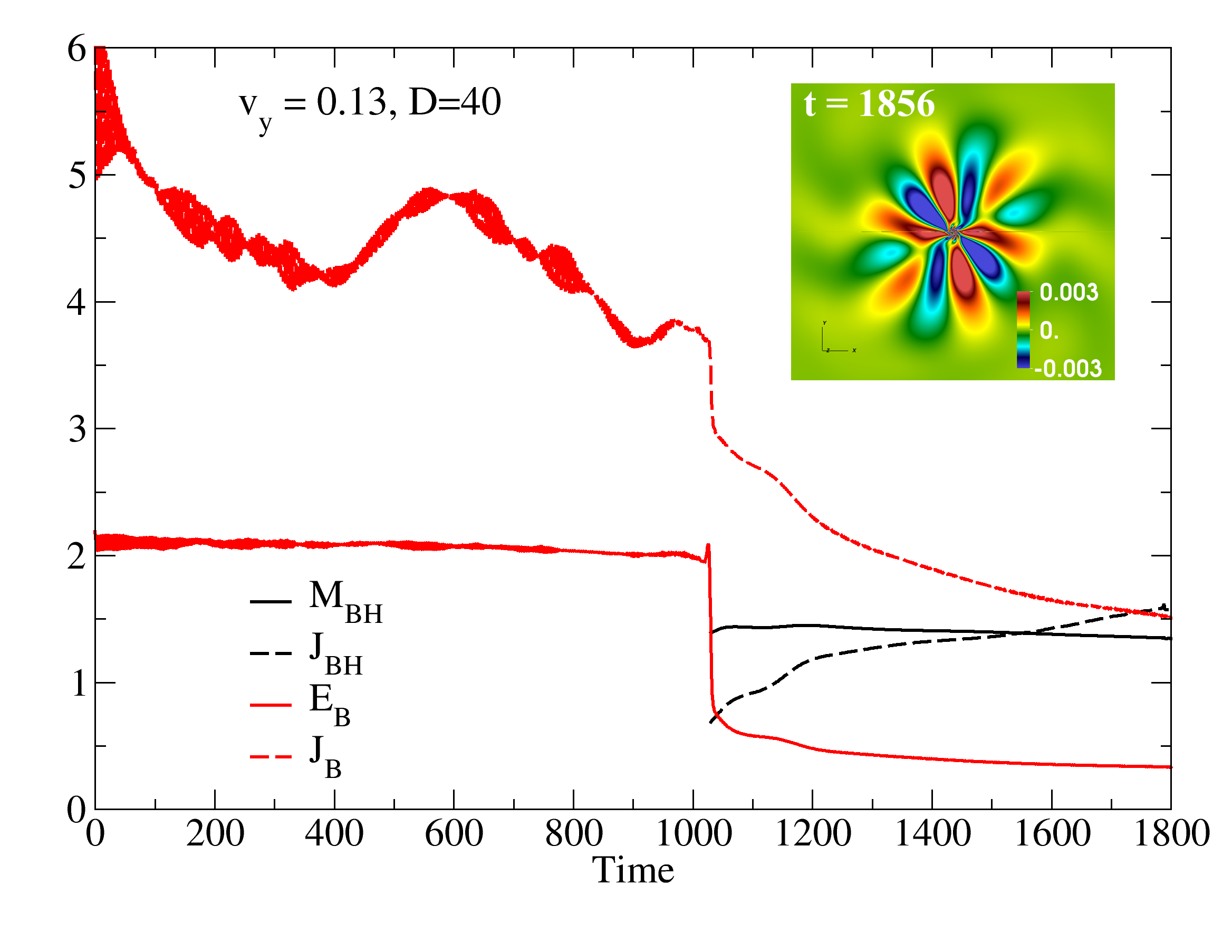}
\caption{Time evolution of $E_{\rm B}, J_{\rm B}, M_{\rm BH}, J_{\rm BH}$ in a simulation with larger $v_y,D$. The inset confirms $m=6$ for the final cloud. 
}
\label{fig6}
\end{figure}

So far, the remnant cloud contains only up to $\sim 0.01 M_i$, $cf.$ Fig.~\ref{fig1} (bottom panel). Significantly higher energy fractions can be obtained by increasing $v_y$. As an illustration, consider a binary of (more compact) vector BSs [$(\omega,M)=(0.91,1.015)$], with $v_y=0.13$ and $D=40$~\footnote{The larger $D$ guarantees a small constraints violation of the initial data.}. The system has $(M_i,J_i)=(2.08,5.98)$ and the BSs perform almost one orbit before merging~\footnote{The initial mass/energy is larger than the mass of the BSs due to the boosts.}. 
For this setup, the Proca  remnant still has $m=6$, but now stores   $\sim 0.15M_i$ and $\sim 0.24J_i$ - Fig.~\ref{fig6}; or, in terms of the final  system, the cloud stores
$\sim18\%$ of the energy and $\sim 50\%$ of the angular momentum.

{\bf {\em Head-on collisions of spinning BSs.}} 
A similar picture occurs for head-on collisions ($i.e.$ with $v_y=0$) of \textit{spinning} vector BSs~\footnote{As mentioned before spinning scalar boson stars are unstable~\cite{Sanchis-Gual:2019ljs}.}. Spinning bosonic stars form a countable number of families, labelled by the azimuthal harmonic index $\bar{m}\in \mathbb{Z}$, which counts the number of (phase) azimuthal nodes. Here we consider mergers of $\bar{m}=1$ and $\bar{m}=2$ spinning vector BSs (static BSs can be seen as the $\bar{m}=0$ case). When the BSs are sufficienly massive, such mergers form a  BH, which is spinning even for head-on collisions. However, the system must possess just enough angular momentum for the final BH to spin up the correct amount as to synchronise with the remnant. This can be achieved by considering  sequences of mergers wherein the frequency of the two (equal mass and parallel spins) initial spinning vector BSs are varied. For this setup, one may envisage varying the frequency of the initial BSs as playing the  same role as varying $v_y$ in the non-spinning BSs mergers discussed before.

\begin{figure}[h!]
\centering
$\bar m=1$ Proca star binary\\
\includegraphics[width=0.30\linewidth]{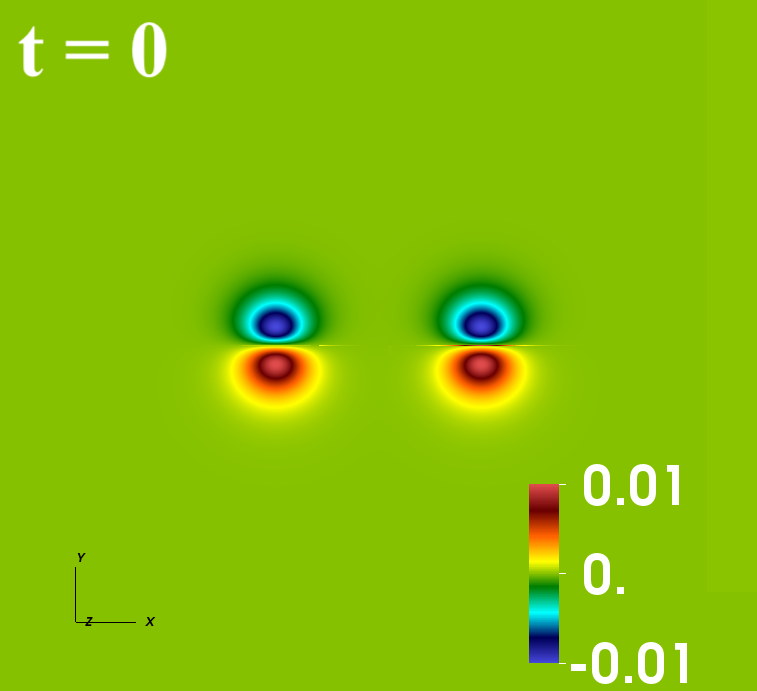}
\includegraphics[width=0.30\linewidth]{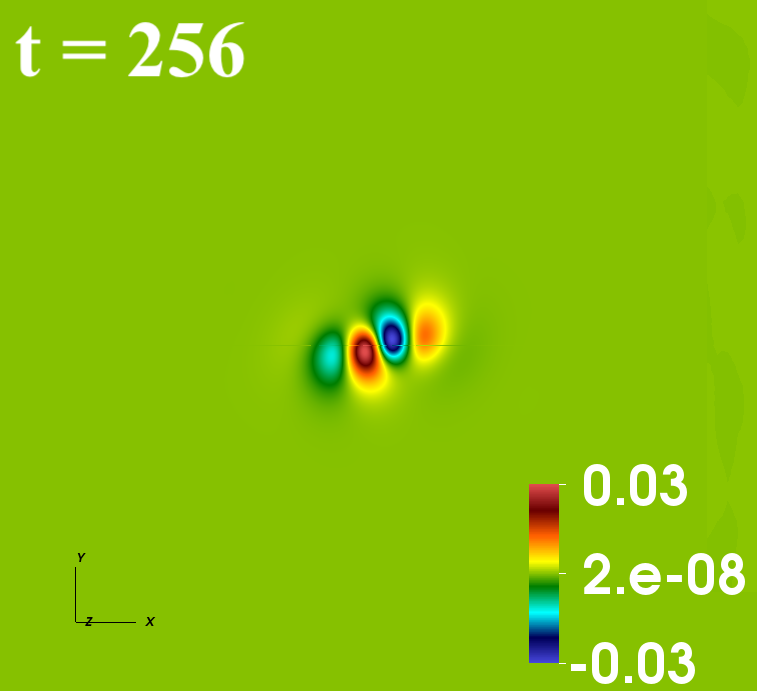}
\includegraphics[width=0.30\linewidth]{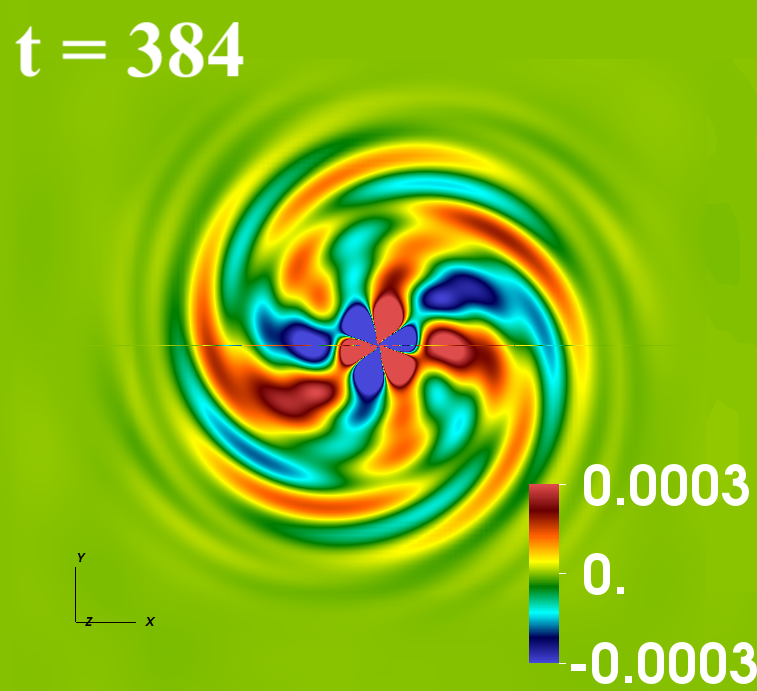}
\includegraphics[width=0.30\linewidth]{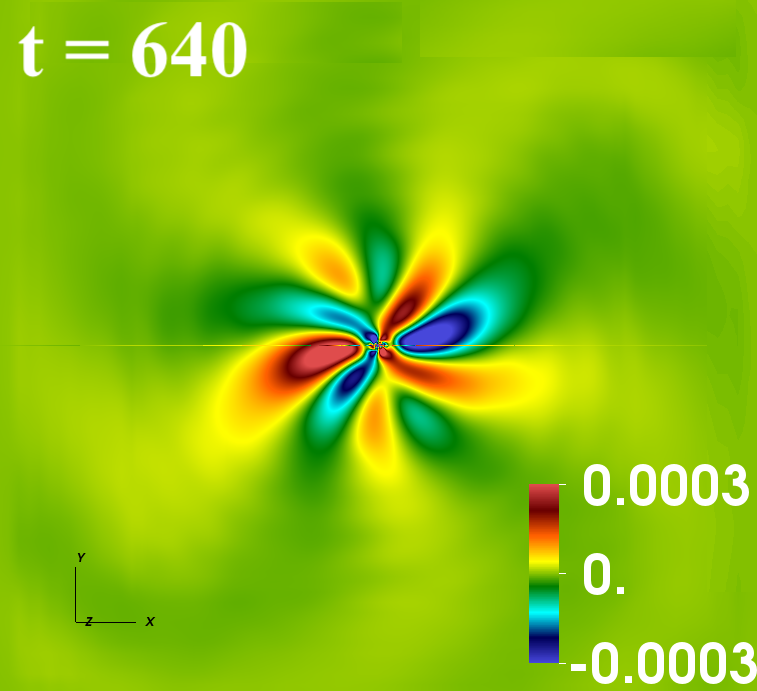}
\includegraphics[width=0.30\linewidth]{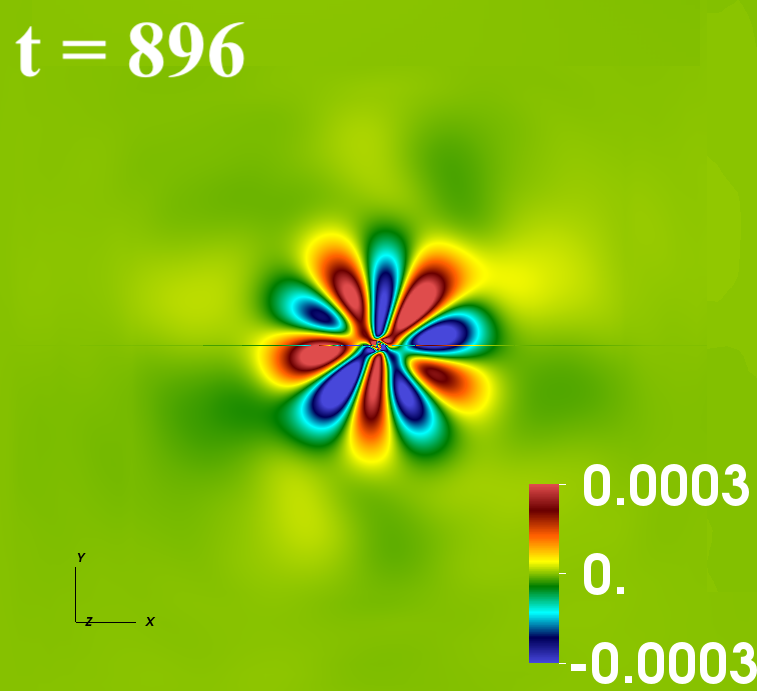}
\includegraphics[width=0.30\linewidth]{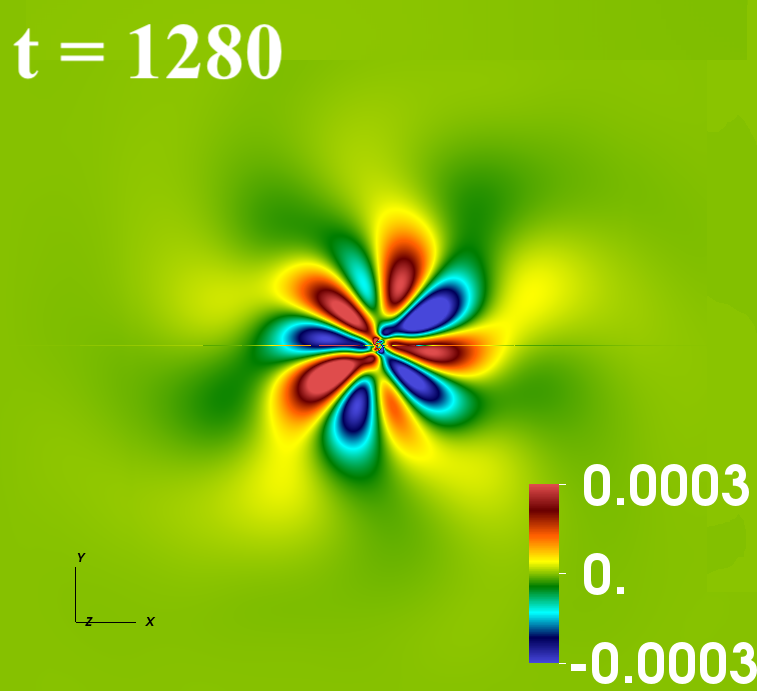}\\
\vspace{0.02\linewidth}
$\bar m=2$ Proca star binary\\
\includegraphics[width=0.30\linewidth]{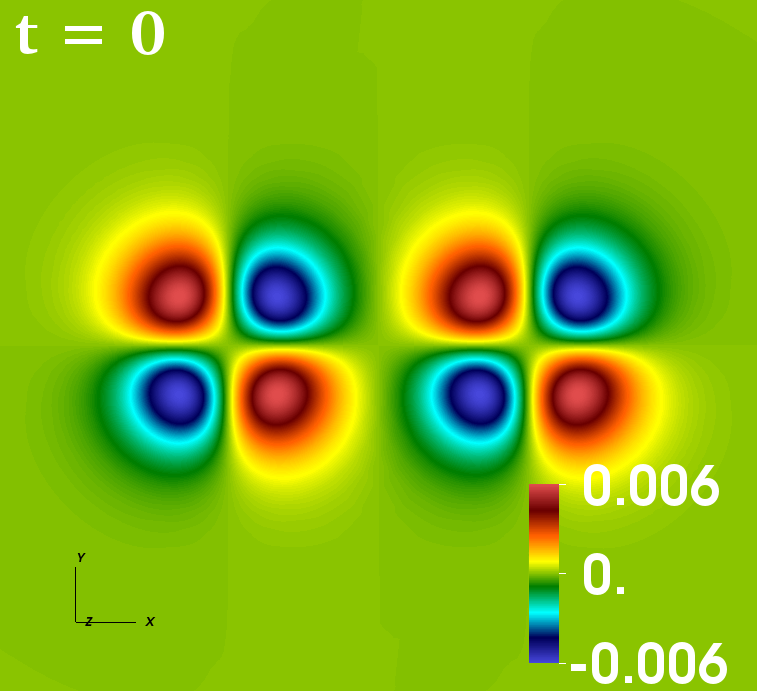}
\includegraphics[width=0.30\linewidth]{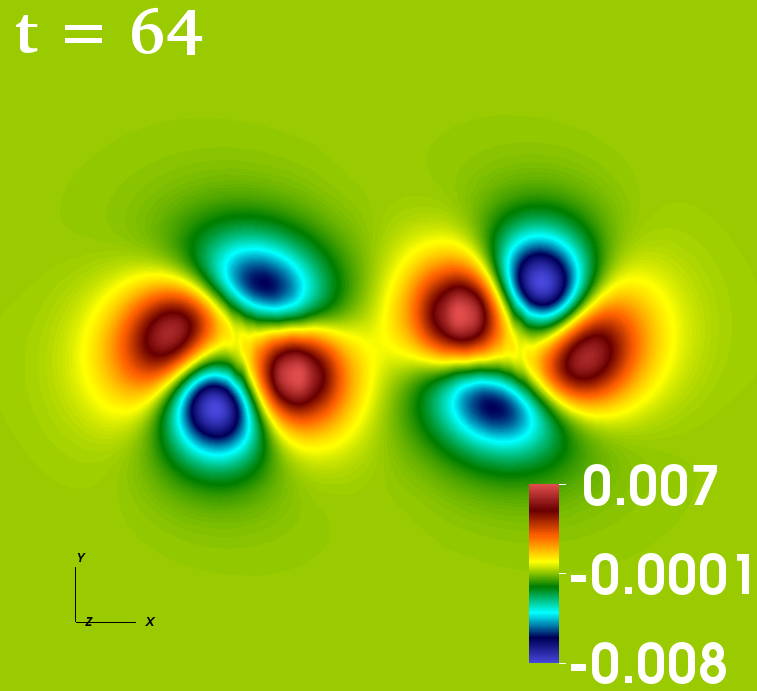}
\includegraphics[width=0.30\linewidth]{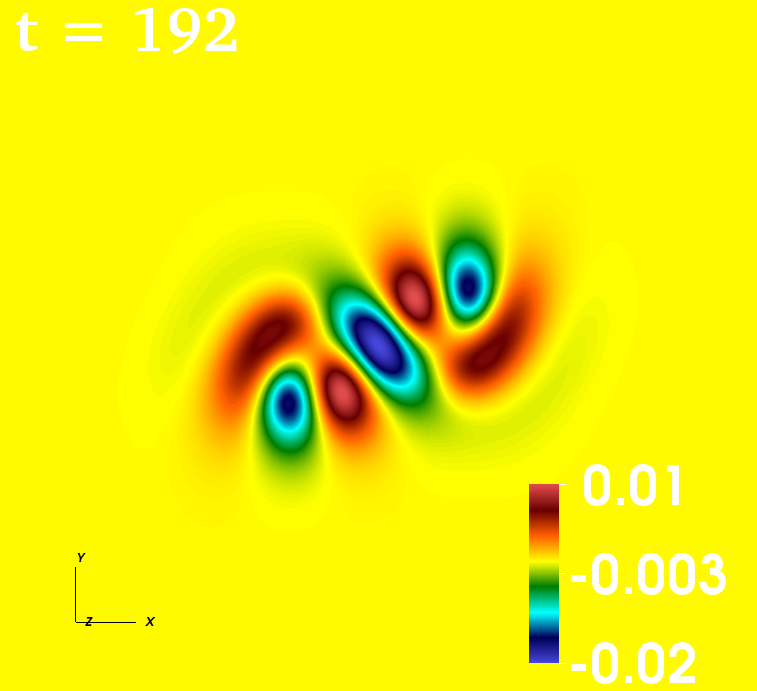}
\includegraphics[width=0.30\linewidth]{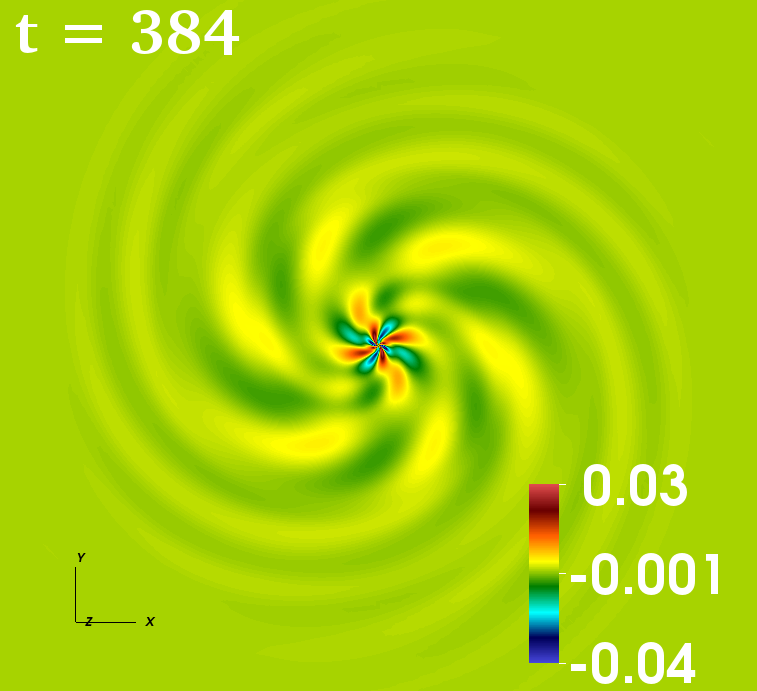}
\includegraphics[width=0.30\linewidth]{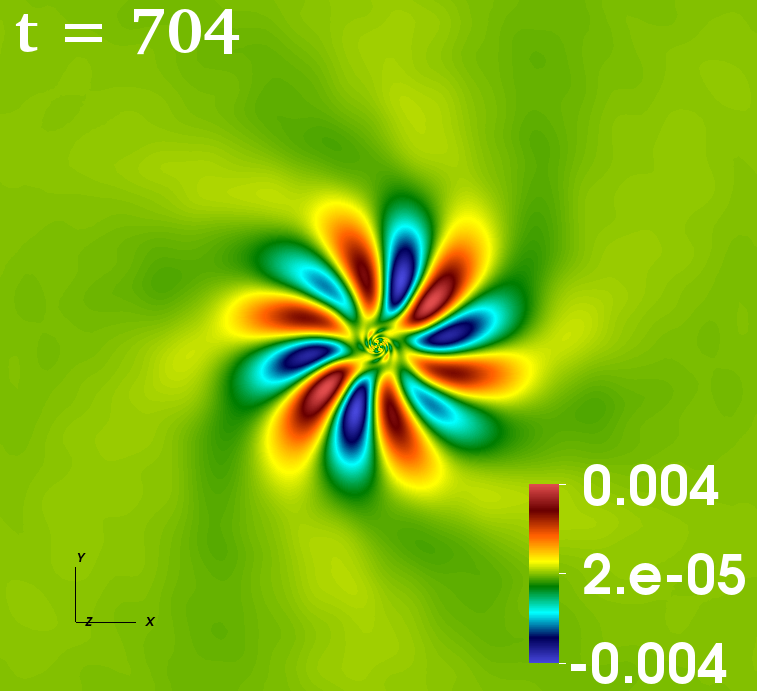}
\includegraphics[width=0.30\linewidth]{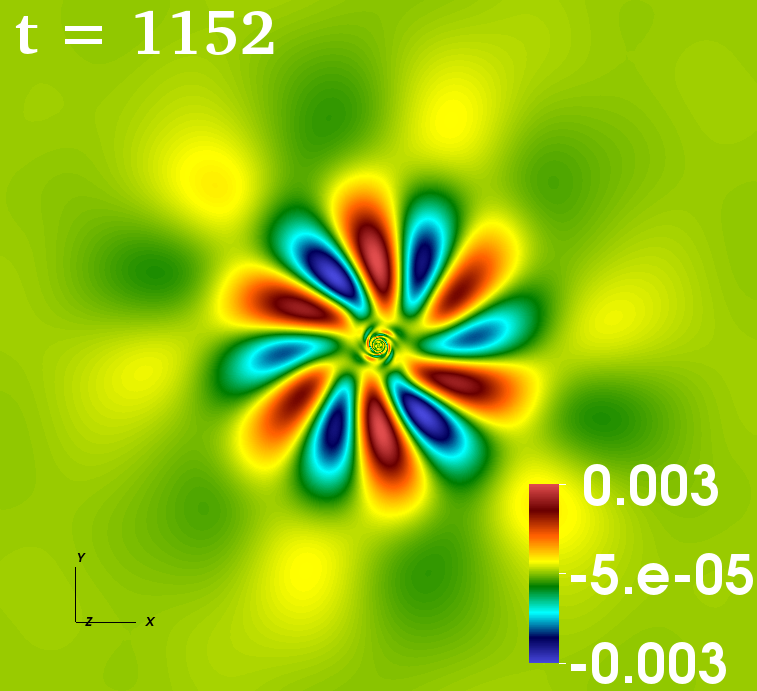}
\caption{Equatorial ($xy$) plane snapshots taken during the time evolution of the mergers of spinning BSs: the two top panels corresponds to a $\bar{m}=1$ model, and the two bottom panels to a $\bar{m}=2$ model. The real part of $\mathcal{X}_{\phi}$ is shown.}
\label{fig7}
\end{figure}

In Fig.~\ref{fig7} we exhibit snapshots of the time evolution of the vector amplitude for two head-on collisions of spinning vector BSs. 
Both collisions have $D=40$, $v_y=0$ and are for systems where synchronisation occurs.  The first collision (top two rows) are for $\bar{m}=1$ stars with $(\omega,M,J)=(0.92, 0.659, 0.677)$. The initial mass and angular momentum of the system are $(M_i, J_i)= (1.33, 1,43)$. An $m=5$ synchronised cloud remains at the end of the simulation, retaining 0.19\% and 0.66\% of $M_i$ and $J_i$, respectively.
The second collision (bottom two rows) are for $\bar{m}=2$ stars with $(\omega,M,J)=(0.93, 1.071, 2.195)$. The initial mass and angular momentum of the system are now $(M_i, J_i)= (2.18, 4.62)$. An $m=6$ synchronised cloud is now obtained at the end of the simulation, retaining 3.2\% and 8.3\% of $M_i$ and $J_i$, respectively.

{\bf {\em SGAs: the BH, the BSs and the mixed channels.}} 
Superradiance of a Kerr BH forms a synchronised bosonic cloud-BH system by \textit{spinning down} the Kerr BH~\cite{east2017superradiant1,Herdeiro:2017phl}. Interestingly, we observe from Fig.~\ref{fig2} that in this new channel,  mergers of BSs lead to a synchronised bosonic cloud-BH system by \textit{spinning up} the BH that results from the merger. This confirms that the synchronised BH-cloud system can be approached from either side.

From the superradiance channel, a universal thermodynamical limit imposes that the bosonic cloud cannot store more than $\sim 29\%$ of the final BH-cloud system~\cite{Brito:2015oca}.  In practice, however, fully non-linear numerical simulations in the Proca case were only able to reach about $9\%$~\cite{east2017superradiant1}. Here, we have shown one example wherein the bosonic cloud stores $\sim18\%$ of the energy and $\sim 50\%$ of the angular momentum of the final BH-cloud system. In fact, there appears to be no theoretical bound from the BSs channel, and we anticipate higher fractions of $M_i,J_i$ can be stored in the bosonic cloud, in particular via processes involving a \textit{mixed} channel.

In order to get synchronisation in a direct, two-step process from the BS channel [the two steps being 1) BH formation, 2) spin up by accretion up to synchronisation], fine tunning of the initial data is required. As illustrated in Fig.~\ref{fig2} a specific value of $v_y$ is necessary.  For smaller $v_y$ there is not enough angular momentum available for $\Omega_H$ to catch up with $\omega/m$ of the cloud; for larger $v_y$, $\Omega_H$ overshoots the synchronisation value, as shown by the simulation with $v_y=0.093$ in Fig.~\ref{fig2} (top panel). In the latter case, the cloud becomes dominated by a superradiant mode. Thus energy and angular momentum extraction from the BH ensue. This is what we call the mixed channel.  We speculate that, in this case, synchronistion is achieved by a three-step process: 1) BH formation, 2) spin up by accretion, 3) superradiant spin down, until synchronisation. Moreover, the results herein indicate the trend that increasing $v_y$, the fractions of $M_B/M_i$ and $J_B/J_i$ increase. Thus, we anticipate that there will be open sets of initial data leading to synchronisation via the three step process. Subsets of these initial data set may well lead to even higher fractions of energy and angular momentum in the final synchronised cloud. Checking this conjecture, however, is challenging. Whereas the two step process occurs in a much shorter time scale than the superradiance evolutions in~\cite{east2017superradiant1}, the three step process will be longer, in particular because high $m$ modes are involved.

Finally, we remark that whereas generic initial data triggering superradiance of Kerr BHs always produces an $m=1$ cloud, corresponding to the fastest growing superradiant mode, generic initial data in the mergers of BSs produce SGAs with different $m$'s, herein illustrated with $m=4,5,6$.

\bigskip


\bigskip

{\bf {\em Acknowledgements.}} 
This work was supported by the Spanish Agencia Estatal de Investigaci\'on (grant PGC2018-095984-B-I00),
by the Generalitat Valenciana (PROMETEO/2019/071 and GRISOLIAP/2019/029), by the Center for Research and Development in Mathematics and Applications (CIDMA) through the Portuguese Foundation for Science and Technology (FCT - Funda\c c\~ao para a Ci\^encia e a Tecnologia), references UIDB/04106/2020 and UIDP/04106/2020, by national funds (OE), through FCT, I.P., in the scope of the framework contract foreseen in the numbers 4, 5 and 6
of the article 23, of the Decree-Law 57/2016, of August 29,
changed by Law 57/2017, of July 19 and by the projects PTDC/FIS-OUT/28407/2017, CERN/FIS-PAR/0027/2019, and UID/FIS/00099/2020 (CENTRA). This work has further been supported by  the  European  Union's  Horizon  2020  research  and  innovation  (RISE) programme H2020-MSCA-RISE-2017 Grant No.~FunFiCO-777740 and by FCT through
Project~No.~UIDB/00099/2020. We would like to acknowledge networking support by the COST Action GWverse CA16104.
MZ also acknowledges financial support provided by FCT/Portugal through the IF
programme, grant IF/00729/2015.
Computations have been performed at the Servei d'Inform\`atica de la Universitat
de Val\`encia, on the ``Baltasar Sete-Sois'' cluster at IST, and at MareNostrum.

\bibliography{num-rel2}

\begin{thebibliography}{45}%
\makeatletter
\providecommand \@ifxundefined [1]{%
 \@ifx{#1\undefined}
}%
\providecommand \@ifnum [1]{%
 \ifnum #1\expandafter \@firstoftwo
 \else \expandafter \@secondoftwo
 \fi
}%
\providecommand \@ifx [1]{%
 \ifx #1\expandafter \@firstoftwo
 \else \expandafter \@secondoftwo
 \fi
}%
\providecommand \natexlab [1]{#1}%
\providecommand \enquote  [1]{``#1''}%
\providecommand \bibnamefont  [1]{#1}%
\providecommand \bibfnamefont [1]{#1}%
\providecommand \citenamefont [1]{#1}%
\providecommand \href@noop [0]{\@secondoftwo}%
\providecommand \href [0]{\begingroup \@sanitize@url \@href}%
\providecommand \@href[1]{\@@startlink{#1}\@@href}%
\providecommand \@@href[1]{\endgroup#1\@@endlink}%
\providecommand \@sanitize@url [0]{\catcode `\\12\catcode `\$12\catcode
  `\&12\catcode `\#12\catcode `\^12\catcode `\_12\catcode `\%12\relax}%
\providecommand \@@startlink[1]{}%
\providecommand \@@endlink[0]{}%
\providecommand \url  [0]{\begingroup\@sanitize@url \@url }%
\providecommand \@url [1]{\endgroup\@href {#1}{\urlprefix }}%
\providecommand \urlprefix  [0]{URL }%
\providecommand \Eprint [0]{\href }%
\providecommand \doibase [0]{http://dx.doi.org/}%
\providecommand \selectlanguage [0]{\@gobble}%
\providecommand \bibinfo  [0]{\@secondoftwo}%
\providecommand \bibfield  [0]{\@secondoftwo}%
\providecommand \translation [1]{[#1]}%
\providecommand \BibitemOpen [0]{}%
\providecommand \bibitemStop [0]{}%
\providecommand \bibitemNoStop [0]{.\EOS\space}%
\providecommand \EOS [0]{\spacefactor3000\relax}%
\providecommand \BibitemShut  [1]{\csname bibitem#1\endcsname}%
\let\auto@bib@innerbib\@empty
\bibitem [{\citenamefont {Mirollo}\ and\ \citenamefont
  {Strogatz}(1990)}]{Mirollo}%
  \BibitemOpen
  \bibfield  {author} {\bibinfo {author} {\bibfnamefont {R.~E.}\ \bibnamefont
  {Mirollo}}\ and\ \bibinfo {author} {\bibfnamefont {S.~H.}\ \bibnamefont
  {Strogatz}},\ }\href {\doibase https://doi.org/10.1137/0150098} {\bibfield
  {journal} {\bibinfo  {journal} {SIAM J. Appl. Math.}\ }\textbf {\bibinfo
  {volume} {50}},\ \bibinfo {pages} {1645} (\bibinfo {year}
  {1990})}\BibitemShut {NoStop}%
\bibitem [{\citenamefont {{Pantaleone}}(2002)}]{2002AmJPh..70..992P}%
  \BibitemOpen
  \bibfield  {author} {\bibinfo {author} {\bibfnamefont {J.}~\bibnamefont
  {{Pantaleone}}},\ }\href {\doibase 10.1119/1.1501118} {\bibfield  {journal}
  {\bibinfo  {journal} {American Journal of Physics}\ }\textbf {\bibinfo
  {volume} {70}},\ \bibinfo {pages} {992} (\bibinfo {year} {2002})}\BibitemShut
  {NoStop}%
\bibitem [{\citenamefont {Strogatz}(2004)}]{strogaatz}%
  \BibitemOpen
  \bibfield  {author} {\bibinfo {author} {\bibfnamefont {S.}~\bibnamefont
  {Strogatz}},\ }\href@noop {} {\emph {\bibinfo {title} {Sync: How Order
  Emerges from Chaos in the Universe, Nature, and Daily Life}}}\ (\bibinfo
  {publisher} {Hachette Books},\ \bibinfo {address} {New York},\ \bibinfo
  {year} {2004})\BibitemShut {NoStop}%
\bibitem [{\citenamefont {{Hut}}(1981)}]{1981A&A....99..126H}%
  \BibitemOpen
  \bibfield  {author} {\bibinfo {author} {\bibfnamefont {P.}~\bibnamefont
  {{Hut}}},\ }\href@noop {} {\bibfield  {journal} {\bibinfo  {journal} {Astron.
  \& Astrophys.}\ }\textbf {\bibinfo {volume} {99}},\ \bibinfo {pages} {126}
  (\bibinfo {year} {1981})}\BibitemShut {NoStop}%
\bibitem [{\citenamefont {{Cheng}}\ \emph {et~al.}(2014)\citenamefont
  {{Cheng}}, \citenamefont {{Lee}},\ and\ \citenamefont
  {{Peale}}}]{2014Icar..233..242C}%
  \BibitemOpen
  \bibfield  {author} {\bibinfo {author} {\bibfnamefont {W.~H.}\ \bibnamefont
  {{Cheng}}}, \bibinfo {author} {\bibfnamefont {M.~H.}\ \bibnamefont {{Lee}}},
  \ and\ \bibinfo {author} {\bibfnamefont {S.~J.}\ \bibnamefont {{Peale}}},\
  }\href {\doibase 10.1016/j.icarus.2014.01.046} {\bibfield  {journal}
  {\bibinfo  {journal} {Icarus}\ }\textbf {\bibinfo {volume} {233}},\ \bibinfo
  {pages} {242} (\bibinfo {year} {2014})},\ \Eprint
  {http://arxiv.org/abs/1402.0625} {arXiv:1402.0625 [astro-ph.EP]} \BibitemShut
  {NoStop}%
\bibitem [{\citenamefont {Brito}\ \emph {et~al.}(2015)\citenamefont {Brito},
  \citenamefont {Cardoso},\ and\ \citenamefont {Pani}}]{Brito:2015oca}%
  \BibitemOpen
  \bibfield  {author} {\bibinfo {author} {\bibfnamefont {R.}~\bibnamefont
  {Brito}}, \bibinfo {author} {\bibfnamefont {V.}~\bibnamefont {Cardoso}}, \
  and\ \bibinfo {author} {\bibfnamefont {P.}~\bibnamefont {Pani}},\ }\href
  {\doibase 10.1007/978-3-319-19000-6} {\bibfield  {journal} {\bibinfo
  {journal} {Lect. Notes Phys.}\ }\textbf {\bibinfo {volume} {906}},\ \bibinfo
  {pages} {pp.1} (\bibinfo {year} {2015})},\ \Eprint
  {http://arxiv.org/abs/1501.06570} {arXiv:1501.06570 [gr-qc]} \BibitemShut
  {NoStop}%
\bibitem [{\citenamefont {East}\ and\ \citenamefont
  {Pretorius}(2017)}]{east2017superradiant1}%
  \BibitemOpen
  \bibfield  {author} {\bibinfo {author} {\bibfnamefont {W.~E.}\ \bibnamefont
  {East}}\ and\ \bibinfo {author} {\bibfnamefont {F.}~\bibnamefont
  {Pretorius}},\ }\href@noop {} {\bibfield  {journal} {\bibinfo  {journal}
  {Physical review letters}\ }\textbf {\bibinfo {volume} {119}},\ \bibinfo
  {pages} {041101} (\bibinfo {year} {2017})}\BibitemShut {NoStop}%
\bibitem [{\citenamefont {Herdeiro}\ and\ \citenamefont
  {Radu}(2017)}]{Herdeiro:2017phl}%
  \BibitemOpen
  \bibfield  {author} {\bibinfo {author} {\bibfnamefont {C.~A.~R.}\
  \bibnamefont {Herdeiro}}\ and\ \bibinfo {author} {\bibfnamefont
  {E.}~\bibnamefont {Radu}},\ }\href {\doibase 10.1103/PhysRevLett.119.261101}
  {\bibfield  {journal} {\bibinfo  {journal} {Phys. Rev. Lett.}\ }\textbf
  {\bibinfo {volume} {119}},\ \bibinfo {pages} {261101} (\bibinfo {year}
  {2017})},\ \Eprint {http://arxiv.org/abs/1706.06597} {arXiv:1706.06597
  [gr-qc]} \BibitemShut {NoStop}%
\bibitem [{\citenamefont {Hui}\ \emph {et~al.}(2017)\citenamefont {Hui},
  \citenamefont {Ostriker}, \citenamefont {Tremaine},\ and\ \citenamefont
  {Witten}}]{Hui:2016ltb}%
  \BibitemOpen
  \bibfield  {author} {\bibinfo {author} {\bibfnamefont {L.}~\bibnamefont
  {Hui}}, \bibinfo {author} {\bibfnamefont {J.~P.}\ \bibnamefont {Ostriker}},
  \bibinfo {author} {\bibfnamefont {S.}~\bibnamefont {Tremaine}}, \ and\
  \bibinfo {author} {\bibfnamefont {E.}~\bibnamefont {Witten}},\ }\href
  {\doibase 10.1103/PhysRevD.95.043541} {\bibfield  {journal} {\bibinfo
  {journal} {Phys. Rev. D}\ }\textbf {\bibinfo {volume} {95}},\ \bibinfo
  {pages} {043541} (\bibinfo {year} {2017})},\ \Eprint
  {http://arxiv.org/abs/1610.08297} {arXiv:1610.08297 [astro-ph.CO]}
  \BibitemShut {NoStop}%
\bibitem [{\citenamefont {Arvanitaki}\ and\ \citenamefont
  {Dubovsky}(2011)}]{Arvanitaki:2010sy}%
  \BibitemOpen
  \bibfield  {author} {\bibinfo {author} {\bibfnamefont {A.}~\bibnamefont
  {Arvanitaki}}\ and\ \bibinfo {author} {\bibfnamefont {S.}~\bibnamefont
  {Dubovsky}},\ }\href {\doibase 10.1103/PhysRevD.83.044026} {\bibfield
  {journal} {\bibinfo  {journal} {Phys.Rev.}\ }\textbf {\bibinfo {volume}
  {D83}},\ \bibinfo {pages} {044026} (\bibinfo {year} {2011})},\ \Eprint
  {http://arxiv.org/abs/1004.3558} {arXiv:1004.3558 [hep-th]} \BibitemShut
  {NoStop}%
\bibitem [{\citenamefont {Baumann}\ \emph {et~al.}(2019)\citenamefont
  {Baumann}, \citenamefont {Chia}, \citenamefont {Stout},\ and\ \citenamefont
  {ter Haar}}]{Baumann:2019eav}%
  \BibitemOpen
  \bibfield  {author} {\bibinfo {author} {\bibfnamefont {D.}~\bibnamefont
  {Baumann}}, \bibinfo {author} {\bibfnamefont {H.~S.}\ \bibnamefont {Chia}},
  \bibinfo {author} {\bibfnamefont {J.}~\bibnamefont {Stout}}, \ and\ \bibinfo
  {author} {\bibfnamefont {L.}~\bibnamefont {ter Haar}},\ }\href {\doibase
  10.1088/1475-7516/2019/12/006} {\bibfield  {journal} {\bibinfo  {journal}
  {JCAP}\ }\textbf {\bibinfo {volume} {12}},\ \bibinfo {pages} {006} (\bibinfo
  {year} {2019})},\ \Eprint {http://arxiv.org/abs/1908.10370} {arXiv:1908.10370
  [gr-qc]} \BibitemShut {NoStop}%
\bibitem [{\citenamefont {Herdeiro}\ and\ \citenamefont
  {Radu}(2014)}]{Herdeiro:2014goa}%
  \BibitemOpen
  \bibfield  {author} {\bibinfo {author} {\bibfnamefont {C.~A.~R.}\
  \bibnamefont {Herdeiro}}\ and\ \bibinfo {author} {\bibfnamefont
  {E.}~\bibnamefont {Radu}},\ }\href {\doibase 10.1103/PhysRevLett.112.221101}
  {\bibfield  {journal} {\bibinfo  {journal} {Phys.Rev.Lett.}\ }\textbf
  {\bibinfo {volume} {112}},\ \bibinfo {pages} {221101} (\bibinfo {year}
  {2014})},\ \Eprint {http://arxiv.org/abs/1403.2757} {arXiv:1403.2757 [gr-qc]}
  \BibitemShut {NoStop}%
\bibitem [{\citenamefont {Herdeiro}\ \emph {et~al.}(2016)\citenamefont
  {Herdeiro}, \citenamefont {Radu},\ and\ \citenamefont
  {Runarsson}}]{Herdeiro:2016tmi}%
  \BibitemOpen
  \bibfield  {author} {\bibinfo {author} {\bibfnamefont {C.}~\bibnamefont
  {Herdeiro}}, \bibinfo {author} {\bibfnamefont {E.}~\bibnamefont {Radu}}, \
  and\ \bibinfo {author} {\bibfnamefont {H.}~\bibnamefont {Runarsson}},\ }\href
  {\doibase 10.1088/0264-9381/33/15/154001} {\bibfield  {journal} {\bibinfo
  {journal} {Class. Quant. Grav.}\ }\textbf {\bibinfo {volume} {33}},\ \bibinfo
  {pages} {154001} (\bibinfo {year} {2016})},\ \Eprint
  {http://arxiv.org/abs/1603.02687} {arXiv:1603.02687 [gr-qc]} \BibitemShut
  {NoStop}%
\bibitem [{\citenamefont {Santos}\ \emph {et~al.}(2020)\citenamefont {Santos},
  \citenamefont {Benone}, \citenamefont {Crispino}, \citenamefont {Herdeiro},\
  and\ \citenamefont {Radu}}]{Santos:2020pmh}%
  \BibitemOpen
  \bibfield  {author} {\bibinfo {author} {\bibfnamefont {N.~M.}\ \bibnamefont
  {Santos}}, \bibinfo {author} {\bibfnamefont {C.~L.}\ \bibnamefont {Benone}},
  \bibinfo {author} {\bibfnamefont {L.~C.}\ \bibnamefont {Crispino}}, \bibinfo
  {author} {\bibfnamefont {C.~A.}\ \bibnamefont {Herdeiro}}, \ and\ \bibinfo
  {author} {\bibfnamefont {E.}~\bibnamefont {Radu}},\ }\href@noop {} {\
  (\bibinfo {year} {2020})},\ \Eprint {http://arxiv.org/abs/2004.09536}
  {arXiv:2004.09536 [gr-qc]} \BibitemShut {NoStop}%
\bibitem [{\citenamefont {Schunck}\ and\ \citenamefont
  {Mielke}(2003)}]{Schunck:2003kk}%
  \BibitemOpen
  \bibfield  {author} {\bibinfo {author} {\bibfnamefont {F.~E.}\ \bibnamefont
  {Schunck}}\ and\ \bibinfo {author} {\bibfnamefont {E.~W.}\ \bibnamefont
  {Mielke}},\ }\href {\doibase 10.1088/0264-9381/20/20/201} {\bibfield
  {journal} {\bibinfo  {journal} {Class. Quant. Grav.}\ }\textbf {\bibinfo
  {volume} {20}},\ \bibinfo {pages} {R301} (\bibinfo {year} {2003})},\ \Eprint
  {http://arxiv.org/abs/0801.0307} {arXiv:0801.0307 [astro-ph]} \BibitemShut
  {NoStop}%
\bibitem [{\citenamefont {Liebling}\ and\ \citenamefont
  {Palenzuela}(2012)}]{Liebling:2012fv}%
  \BibitemOpen
  \bibfield  {author} {\bibinfo {author} {\bibfnamefont {S.~L.}\ \bibnamefont
  {Liebling}}\ and\ \bibinfo {author} {\bibfnamefont {C.}~\bibnamefont
  {Palenzuela}},\ }\href {\doibase 10.12942/lrr-2012-6} {\bibfield  {journal}
  {\bibinfo  {journal} {Living Rev. Rel.}\ }\textbf {\bibinfo {volume} {15}},\
  \bibinfo {pages} {6} (\bibinfo {year} {2012})},\ \Eprint
  {http://arxiv.org/abs/1202.5809} {arXiv:1202.5809 [gr-qc]} \BibitemShut
  {NoStop}%
\bibitem [{\citenamefont {Brito}\ \emph {et~al.}(2016)\citenamefont {Brito},
  \citenamefont {Cardoso}, \citenamefont {Herdeiro},\ and\ \citenamefont
  {Radu}}]{Brito:2015pxa}%
  \BibitemOpen
  \bibfield  {author} {\bibinfo {author} {\bibfnamefont {R.}~\bibnamefont
  {Brito}}, \bibinfo {author} {\bibfnamefont {V.}~\bibnamefont {Cardoso}},
  \bibinfo {author} {\bibfnamefont {C.~A.~R.}\ \bibnamefont {Herdeiro}}, \ and\
  \bibinfo {author} {\bibfnamefont {E.}~\bibnamefont {Radu}},\ }\href {\doibase
  10.1016/j.physletb.2015.11.051} {\bibfield  {journal} {\bibinfo  {journal}
  {Phys. Lett.}\ }\textbf {\bibinfo {volume} {B752}},\ \bibinfo {pages} {291}
  (\bibinfo {year} {2016})},\ \Eprint {http://arxiv.org/abs/1508.05395}
  {arXiv:1508.05395 [gr-qc]} \BibitemShut {NoStop}%
\bibitem [{\citenamefont {Herdeiro}\ \emph {et~al.}(2017)\citenamefont
  {Herdeiro}, \citenamefont {Pombo},\ and\ \citenamefont
  {Radu}}]{Herdeiro:2017fhv}%
  \BibitemOpen
  \bibfield  {author} {\bibinfo {author} {\bibfnamefont {C.~A.~R.}\
  \bibnamefont {Herdeiro}}, \bibinfo {author} {\bibfnamefont {A.~M.}\
  \bibnamefont {Pombo}}, \ and\ \bibinfo {author} {\bibfnamefont
  {E.}~\bibnamefont {Radu}},\ }\href {\doibase 10.1016/j.physletb.2017.09.036}
  {\bibfield  {journal} {\bibinfo  {journal} {Phys. Lett. B}\ }\textbf
  {\bibinfo {volume} {773}},\ \bibinfo {pages} {654} (\bibinfo {year}
  {2017})},\ \Eprint {http://arxiv.org/abs/1708.05674} {arXiv:1708.05674
  [gr-qc]} \BibitemShut {NoStop}%
\bibitem [{\citenamefont {Herdeiro}\ \emph {et~al.}(2019)\citenamefont
  {Herdeiro}, \citenamefont {Perapechka}, \citenamefont {Radu},\ and\
  \citenamefont {Shnir}}]{Herdeiro:2019mbz}%
  \BibitemOpen
  \bibfield  {author} {\bibinfo {author} {\bibfnamefont {C.}~\bibnamefont
  {Herdeiro}}, \bibinfo {author} {\bibfnamefont {I.}~\bibnamefont
  {Perapechka}}, \bibinfo {author} {\bibfnamefont {E.}~\bibnamefont {Radu}}, \
  and\ \bibinfo {author} {\bibfnamefont {Y.}~\bibnamefont {Shnir}},\ }\href
  {\doibase 10.1016/j.physletb.2019.134845} {\bibfield  {journal} {\bibinfo
  {journal} {Phys. Lett. B}\ }\textbf {\bibinfo {volume} {797}},\ \bibinfo
  {pages} {134845} (\bibinfo {year} {2019})},\ \Eprint
  {http://arxiv.org/abs/1906.05386} {arXiv:1906.05386 [gr-qc]} \BibitemShut
  {NoStop}%
\bibitem [{\citenamefont {Seidel}\ and\ \citenamefont
  {Suen}(1990)}]{Seidel:1990jh}%
  \BibitemOpen
  \bibfield  {author} {\bibinfo {author} {\bibfnamefont {E.}~\bibnamefont
  {Seidel}}\ and\ \bibinfo {author} {\bibfnamefont {W.-M.}\ \bibnamefont
  {Suen}},\ }\href {\doibase 10.1103/PhysRevD.42.384} {\bibfield  {journal}
  {\bibinfo  {journal} {Phys. Rev.}\ }\textbf {\bibinfo {volume} {D42}},\
  \bibinfo {pages} {384} (\bibinfo {year} {1990})}\BibitemShut {NoStop}%
\bibitem [{\citenamefont {Seidel}\ and\ \citenamefont
  {Suen}(1994)}]{seidel1994formation}%
  \BibitemOpen
  \bibfield  {author} {\bibinfo {author} {\bibfnamefont {E.}~\bibnamefont
  {Seidel}}\ and\ \bibinfo {author} {\bibfnamefont {W.-M.}\ \bibnamefont
  {Suen}},\ }\href@noop {} {\bibfield  {journal} {\bibinfo  {journal} {Physical
  review letters}\ }\textbf {\bibinfo {volume} {72}},\ \bibinfo {pages} {2516}
  (\bibinfo {year} {1994})}\BibitemShut {NoStop}%
\bibitem [{\citenamefont {Sanchis-Gual}\ \emph {et~al.}(2017)\citenamefont
  {Sanchis-Gual}, \citenamefont {Herdeiro}, \citenamefont {Radu}, \citenamefont
  {Degollado},\ and\ \citenamefont {Font}}]{sanchis2017numerical}%
  \BibitemOpen
  \bibfield  {author} {\bibinfo {author} {\bibfnamefont {N.}~\bibnamefont
  {Sanchis-Gual}}, \bibinfo {author} {\bibfnamefont {C.}~\bibnamefont
  {Herdeiro}}, \bibinfo {author} {\bibfnamefont {E.}~\bibnamefont {Radu}},
  \bibinfo {author} {\bibfnamefont {J.~C.}\ \bibnamefont {Degollado}}, \ and\
  \bibinfo {author} {\bibfnamefont {J.~A.}\ \bibnamefont {Font}},\ }\href@noop
  {} {\bibfield  {journal} {\bibinfo  {journal} {Phys. Rev. D}\ }\textbf
  {\bibinfo {volume} {95}},\ \bibinfo {pages} {104028} (\bibinfo {year}
  {2017})}\BibitemShut {NoStop}%
\bibitem [{\citenamefont {Di~Giovanni}\ \emph {et~al.}(2018)\citenamefont
  {Di~Giovanni}, \citenamefont {Sanchis-Gual}, \citenamefont {Herdeiro},\ and\
  \citenamefont {Font}}]{di2018dynamical}%
  \BibitemOpen
  \bibfield  {author} {\bibinfo {author} {\bibfnamefont {F.}~\bibnamefont
  {Di~Giovanni}}, \bibinfo {author} {\bibfnamefont {N.}~\bibnamefont
  {Sanchis-Gual}}, \bibinfo {author} {\bibfnamefont {C.~A.~R.}\ \bibnamefont
  {Herdeiro}}, \ and\ \bibinfo {author} {\bibfnamefont {J.~A.}\ \bibnamefont
  {Font}},\ }\href {\doibase 10.1103/PhysRevD.98.064044} {\bibfield  {journal}
  {\bibinfo  {journal} {Phys. Rev.}\ }\textbf {\bibinfo {volume} {D98}},\
  \bibinfo {pages} {064044} (\bibinfo {year} {2018})},\ \Eprint
  {http://arxiv.org/abs/1803.04802} {arXiv:1803.04802 [gr-qc]} \BibitemShut
  {NoStop}%
\bibitem [{\citenamefont {Palenzuela}\ \emph {et~al.}(2008)\citenamefont
  {Palenzuela}, \citenamefont {Lehner},\ and\ \citenamefont
  {Liebling}}]{palenzuela2008orbital}%
  \BibitemOpen
  \bibfield  {author} {\bibinfo {author} {\bibfnamefont {C.}~\bibnamefont
  {Palenzuela}}, \bibinfo {author} {\bibfnamefont {L.}~\bibnamefont {Lehner}},
  \ and\ \bibinfo {author} {\bibfnamefont {S.~L.}\ \bibnamefont {Liebling}},\
  }\href@noop {} {\bibfield  {journal} {\bibinfo  {journal} {Physical Review
  D}\ }\textbf {\bibinfo {volume} {77}},\ \bibinfo {pages} {044036} (\bibinfo
  {year} {2008})}\BibitemShut {NoStop}%
\bibitem [{\citenamefont {Cardoso}\ \emph {et~al.}(2016)\citenamefont
  {Cardoso}, \citenamefont {Hopper}, \citenamefont {Macedo}, \citenamefont
  {Palenzuela},\ and\ \citenamefont {Pani}}]{cardoso2016gravitational}%
  \BibitemOpen
  \bibfield  {author} {\bibinfo {author} {\bibfnamefont {V.}~\bibnamefont
  {Cardoso}}, \bibinfo {author} {\bibfnamefont {S.}~\bibnamefont {Hopper}},
  \bibinfo {author} {\bibfnamefont {C.~F.}\ \bibnamefont {Macedo}}, \bibinfo
  {author} {\bibfnamefont {C.}~\bibnamefont {Palenzuela}}, \ and\ \bibinfo
  {author} {\bibfnamefont {P.}~\bibnamefont {Pani}},\ }\href@noop {} {\bibfield
   {journal} {\bibinfo  {journal} {Physical review D}\ }\textbf {\bibinfo
  {volume} {94}},\ \bibinfo {pages} {084031} (\bibinfo {year}
  {2016})}\BibitemShut {NoStop}%
\bibitem [{\citenamefont {Bezares}\ \emph {et~al.}(2017)\citenamefont
  {Bezares}, \citenamefont {Palenzuela},\ and\ \citenamefont
  {Bona}}]{bezares2017final}%
  \BibitemOpen
  \bibfield  {author} {\bibinfo {author} {\bibfnamefont {M.}~\bibnamefont
  {Bezares}}, \bibinfo {author} {\bibfnamefont {C.}~\bibnamefont {Palenzuela}},
  \ and\ \bibinfo {author} {\bibfnamefont {C.}~\bibnamefont {Bona}},\
  }\href@noop {} {\bibfield  {journal} {\bibinfo  {journal} {Physical Review
  D}\ }\textbf {\bibinfo {volume} {95}},\ \bibinfo {pages} {124005} (\bibinfo
  {year} {2017})}\BibitemShut {NoStop}%
\bibitem [{\citenamefont {Sanchis-Gual}\ \emph
  {et~al.}(2019{\natexlab{a}})\citenamefont {Sanchis-Gual}, \citenamefont
  {Herdeiro}, \citenamefont {Font}, \citenamefont {Radu},\ and\ \citenamefont
  {Di~Giovanni}}]{sanchis2019head}%
  \BibitemOpen
  \bibfield  {author} {\bibinfo {author} {\bibfnamefont {N.}~\bibnamefont
  {Sanchis-Gual}}, \bibinfo {author} {\bibfnamefont {C.}~\bibnamefont
  {Herdeiro}}, \bibinfo {author} {\bibfnamefont {J.~A.}\ \bibnamefont {Font}},
  \bibinfo {author} {\bibfnamefont {E.}~\bibnamefont {Radu}}, \ and\ \bibinfo
  {author} {\bibfnamefont {F.}~\bibnamefont {Di~Giovanni}},\ }\href@noop {}
  {\bibfield  {journal} {\bibinfo  {journal} {Phys. Rev. D}\ }\textbf {\bibinfo
  {volume} {99}},\ \bibinfo {pages} {024017} (\bibinfo {year}
  {2019}{\natexlab{a}})}\BibitemShut {NoStop}%
\bibitem [{\citenamefont {Sanchis-Gual}\ \emph
  {et~al.}(2019{\natexlab{b}})\citenamefont {Sanchis-Gual}, \citenamefont
  {Di~Giovanni}, \citenamefont {Zilhão}, \citenamefont {Herdeiro},
  \citenamefont {Cerdá-Durán}, \citenamefont {Font},\ and\ \citenamefont
  {Radu}}]{Sanchis-Gual:2019ljs}%
  \BibitemOpen
  \bibfield  {author} {\bibinfo {author} {\bibfnamefont {N.}~\bibnamefont
  {Sanchis-Gual}}, \bibinfo {author} {\bibfnamefont {F.}~\bibnamefont
  {Di~Giovanni}}, \bibinfo {author} {\bibfnamefont {M.}~\bibnamefont
  {Zilhão}}, \bibinfo {author} {\bibfnamefont {C.}~\bibnamefont {Herdeiro}},
  \bibinfo {author} {\bibfnamefont {P.}~\bibnamefont {Cerdá-Durán}}, \bibinfo
  {author} {\bibfnamefont {J.}~\bibnamefont {Font}}, \ and\ \bibinfo {author}
  {\bibfnamefont {E.}~\bibnamefont {Radu}},\ }\href {\doibase
  10.1103/PhysRevLett.123.221101} {\bibfield  {journal} {\bibinfo  {journal}
  {Phys. Rev. Lett.}\ }\textbf {\bibinfo {volume} {123}},\ \bibinfo {pages}
  {221101} (\bibinfo {year} {2019}{\natexlab{b}})},\ \Eprint
  {http://arxiv.org/abs/1907.12565} {arXiv:1907.12565 [gr-qc]} \BibitemShut
  {NoStop}%
\bibitem [{\citenamefont {Cunha}\ \emph {et~al.}(2017)\citenamefont {Cunha},
  \citenamefont {Font}, \citenamefont {Herdeiro}, \citenamefont {Radu},
  \citenamefont {Sanchis-Gual},\ and\ \citenamefont {Zilhão}}]{Cunha:2017wao}%
  \BibitemOpen
  \bibfield  {author} {\bibinfo {author} {\bibfnamefont {P.~V.~P.}\
  \bibnamefont {Cunha}}, \bibinfo {author} {\bibfnamefont {J.~A.}\ \bibnamefont
  {Font}}, \bibinfo {author} {\bibfnamefont {C.}~\bibnamefont {Herdeiro}},
  \bibinfo {author} {\bibfnamefont {E.}~\bibnamefont {Radu}}, \bibinfo {author}
  {\bibfnamefont {N.}~\bibnamefont {Sanchis-Gual}}, \ and\ \bibinfo {author}
  {\bibfnamefont {M.}~\bibnamefont {Zilhão}},\ }\href {\doibase
  10.1103/PhysRevD.96.104040} {\bibfield  {journal} {\bibinfo  {journal} {Phys.
  Rev.}\ }\textbf {\bibinfo {volume} {D96}},\ \bibinfo {pages} {104040}
  (\bibinfo {year} {2017})},\ \Eprint {http://arxiv.org/abs/1709.06118}
  {arXiv:1709.06118 [gr-qc]} \BibitemShut {NoStop}%
\bibitem [{\citenamefont {Zilh\~ao}\ \emph {et~al.}(2015)\citenamefont
  {Zilh\~ao}, \citenamefont {Witek},\ and\ \citenamefont
  {Cardoso}}]{Zilhao:2015tya}%
  \BibitemOpen
  \bibfield  {author} {\bibinfo {author} {\bibfnamefont {M.}~\bibnamefont
  {Zilh\~ao}}, \bibinfo {author} {\bibfnamefont {H.}~\bibnamefont {Witek}}, \
  and\ \bibinfo {author} {\bibfnamefont {V.}~\bibnamefont {Cardoso}},\ }\href
  {\doibase 10.1088/0264-9381/32/23/234003} {\bibfield  {journal} {\bibinfo
  {journal} {Class. Quant. Grav.}\ }\textbf {\bibinfo {volume} {32}},\ \bibinfo
  {pages} {234003} (\bibinfo {year} {2015})},\ \Eprint
  {http://arxiv.org/abs/1505.00797} {arXiv:1505.00797 [gr-qc]} \BibitemShut
  {NoStop}%
\bibitem [{\citenamefont {Loffler}\ \emph {et~al.}(2012)\citenamefont
  {Loffler}, \citenamefont {Faber}, \citenamefont {Bentivegna}, \citenamefont
  {Bode}, \citenamefont {Diener} \emph {et~al.}}]{Loffler:2011ay}%
  \BibitemOpen
  \bibfield  {author} {\bibinfo {author} {\bibfnamefont {F.}~\bibnamefont
  {Loffler}}, \bibinfo {author} {\bibfnamefont {J.}~\bibnamefont {Faber}},
  \bibinfo {author} {\bibfnamefont {E.}~\bibnamefont {Bentivegna}}, \bibinfo
  {author} {\bibfnamefont {T.}~\bibnamefont {Bode}}, \bibinfo {author}
  {\bibfnamefont {P.}~\bibnamefont {Diener}},  \emph {et~al.},\ }\href
  {\doibase 10.1088/0264-9381/29/11/115001} {\bibfield  {journal} {\bibinfo
  {journal} {Class.Quant.Grav.}\ }\textbf {\bibinfo {volume} {29}},\ \bibinfo
  {pages} {115001} (\bibinfo {year} {2012})},\ \Eprint
  {http://arxiv.org/abs/1111.3344} {arXiv:1111.3344 [gr-qc]} \BibitemShut
  {NoStop}%
\bibitem [{\citenamefont {Zilh{\~{a}}o}\ and\ \citenamefont
  {L{\"{o}}ffler}(2013)}]{Zilhao:2013hia}%
  \BibitemOpen
  \bibfield  {author} {\bibinfo {author} {\bibfnamefont {M.}~\bibnamefont
  {Zilh{\~{a}}o}}\ and\ \bibinfo {author} {\bibfnamefont {F.}~\bibnamefont
  {L{\"{o}}ffler}},\ }\href {\doibase 10.1142/S0217751X13400149} {\bibfield
  {journal} {\bibinfo  {journal} {Int.J.Mod.Phys.}\ }\textbf {\bibinfo {volume}
  {A28}},\ \bibinfo {pages} {1340014} (\bibinfo {year} {2013})},\ \Eprint
  {http://arxiv.org/abs/1305.5299} {arXiv:1305.5299 [gr-qc]} \BibitemShut
  {NoStop}%
\bibitem [{\citenamefont {Babiuc-Hamilton}\ \emph {et~al.}(2019)\citenamefont
  {Babiuc-Hamilton}, \citenamefont {Brandt}, \citenamefont {Diener},
  \citenamefont {Elley}, \citenamefont {Etienne}, \citenamefont {Ficarra},
  \citenamefont {Haas}, \citenamefont {Witek}, \citenamefont {Alcubierre},
  \citenamefont {Alic}, \citenamefont {Allen}, \citenamefont {Ansorg},
  \citenamefont {Baiotti}, \citenamefont {Benger}, \citenamefont {Bentivegna},
  \citenamefont {Bernuzzi}, \citenamefont {Bode}, \citenamefont {Bruegmann},
  \citenamefont {Corvino}, \citenamefont {De~Pietri}, \citenamefont
  {Dimmelmeier}, \citenamefont {Dooley}, \citenamefont {Dorband}, \citenamefont
  {El~Khamra}, \citenamefont {Faber}, \citenamefont {Font}, \citenamefont
  {Frieben}, \citenamefont {Giacomazzo}, \citenamefont {Goodale}, \citenamefont
  {Gundlach}, \citenamefont {Hawke}, \citenamefont {Hawley}, \citenamefont
  {Hinder}, \citenamefont {Husa}, \citenamefont {Iyer}, \citenamefont
  {Kellermann}, \citenamefont {Knapp}, \citenamefont {Koppitz}, \citenamefont
  {Lanferman}, \citenamefont {Löffler}, \citenamefont {Masso}, \citenamefont
  {Menger}, \citenamefont {Merzky}, \citenamefont {Miller}, \citenamefont
  {Moesta}, \citenamefont {Montero}, \citenamefont {Mundim}, \citenamefont
  {Nerozzi}, \citenamefont {Ott}, \citenamefont {Paruchuri}, \citenamefont
  {Pollney}, \citenamefont {Radice}, \citenamefont {Radke}, \citenamefont
  {Reisswig}, \citenamefont {Rezzolla}, \citenamefont {Rideout}, \citenamefont
  {Ripeanu}, \citenamefont {Schnetter}, \citenamefont {Schutz}, \citenamefont
  {Seidel}, \citenamefont {Seidel}, \citenamefont {Shalf}, \citenamefont
  {Sperhake}, \citenamefont {Stergioulas}, \citenamefont {Suen}, \citenamefont
  {Szilagyi}, \citenamefont {Takahashi}, \citenamefont {Thomas}, \citenamefont
  {Thornburg}, \citenamefont {Tobias}, \citenamefont {Tonita}, \citenamefont
  {Walker}, \citenamefont {Wan}, \citenamefont {Wardell}, \citenamefont
  {Zilhão}, \citenamefont {Zink},\ and\ \citenamefont
  {Zlochower}}]{EinsteinToolkit:2019_10}%
  \BibitemOpen
  \bibfield  {author} {\bibinfo {author} {\bibfnamefont {M.}~\bibnamefont
  {Babiuc-Hamilton}}, \bibinfo {author} {\bibfnamefont {S.~R.}\ \bibnamefont
  {Brandt}}, \bibinfo {author} {\bibfnamefont {P.}~\bibnamefont {Diener}},
  \bibinfo {author} {\bibfnamefont {M.}~\bibnamefont {Elley}}, \bibinfo
  {author} {\bibfnamefont {Z.}~\bibnamefont {Etienne}}, \bibinfo {author}
  {\bibfnamefont {G.}~\bibnamefont {Ficarra}}, \bibinfo {author} {\bibfnamefont
  {R.}~\bibnamefont {Haas}}, \bibinfo {author} {\bibfnamefont {H.}~\bibnamefont
  {Witek}}, \bibinfo {author} {\bibfnamefont {M.}~\bibnamefont {Alcubierre}},
  \bibinfo {author} {\bibfnamefont {D.}~\bibnamefont {Alic}}, \bibinfo {author}
  {\bibfnamefont {G.}~\bibnamefont {Allen}}, \bibinfo {author} {\bibfnamefont
  {M.}~\bibnamefont {Ansorg}}, \bibinfo {author} {\bibfnamefont
  {L.}~\bibnamefont {Baiotti}}, \bibinfo {author} {\bibfnamefont
  {W.}~\bibnamefont {Benger}}, \bibinfo {author} {\bibfnamefont
  {E.}~\bibnamefont {Bentivegna}}, \bibinfo {author} {\bibfnamefont
  {S.}~\bibnamefont {Bernuzzi}}, \bibinfo {author} {\bibfnamefont
  {T.}~\bibnamefont {Bode}}, \bibinfo {author} {\bibfnamefont {B.}~\bibnamefont
  {Bruegmann}}, \bibinfo {author} {\bibfnamefont {G.}~\bibnamefont {Corvino}},
  \bibinfo {author} {\bibfnamefont {R.}~\bibnamefont {De~Pietri}}, \bibinfo
  {author} {\bibfnamefont {H.}~\bibnamefont {Dimmelmeier}}, \bibinfo {author}
  {\bibfnamefont {R.}~\bibnamefont {Dooley}}, \bibinfo {author} {\bibfnamefont
  {N.}~\bibnamefont {Dorband}}, \bibinfo {author} {\bibfnamefont
  {Y.}~\bibnamefont {El~Khamra}}, \bibinfo {author} {\bibfnamefont
  {J.}~\bibnamefont {Faber}}, \bibinfo {author} {\bibfnamefont
  {T.}~\bibnamefont {Font}}, \bibinfo {author} {\bibfnamefont {J.}~\bibnamefont
  {Frieben}}, \bibinfo {author} {\bibfnamefont {B.}~\bibnamefont {Giacomazzo}},
  \bibinfo {author} {\bibfnamefont {T.}~\bibnamefont {Goodale}}, \bibinfo
  {author} {\bibfnamefont {C.}~\bibnamefont {Gundlach}}, \bibinfo {author}
  {\bibfnamefont {I.}~\bibnamefont {Hawke}}, \bibinfo {author} {\bibfnamefont
  {S.}~\bibnamefont {Hawley}}, \bibinfo {author} {\bibfnamefont
  {I.}~\bibnamefont {Hinder}}, \bibinfo {author} {\bibfnamefont
  {S.}~\bibnamefont {Husa}}, \bibinfo {author} {\bibfnamefont {S.}~\bibnamefont
  {Iyer}}, \bibinfo {author} {\bibfnamefont {T.}~\bibnamefont {Kellermann}},
  \bibinfo {author} {\bibfnamefont {A.}~\bibnamefont {Knapp}}, \bibinfo
  {author} {\bibfnamefont {M.}~\bibnamefont {Koppitz}}, \bibinfo {author}
  {\bibfnamefont {G.}~\bibnamefont {Lanferman}}, \bibinfo {author}
  {\bibfnamefont {F.}~\bibnamefont {Löffler}}, \bibinfo {author}
  {\bibfnamefont {J.}~\bibnamefont {Masso}}, \bibinfo {author} {\bibfnamefont
  {L.}~\bibnamefont {Menger}}, \bibinfo {author} {\bibfnamefont
  {A.}~\bibnamefont {Merzky}}, \bibinfo {author} {\bibfnamefont
  {M.}~\bibnamefont {Miller}}, \bibinfo {author} {\bibfnamefont
  {P.}~\bibnamefont {Moesta}}, \bibinfo {author} {\bibfnamefont
  {P.}~\bibnamefont {Montero}}, \bibinfo {author} {\bibfnamefont
  {B.}~\bibnamefont {Mundim}}, \bibinfo {author} {\bibfnamefont
  {A.}~\bibnamefont {Nerozzi}}, \bibinfo {author} {\bibfnamefont
  {C.}~\bibnamefont {Ott}}, \bibinfo {author} {\bibfnamefont {R.}~\bibnamefont
  {Paruchuri}}, \bibinfo {author} {\bibfnamefont {D.}~\bibnamefont {Pollney}},
  \bibinfo {author} {\bibfnamefont {D.}~\bibnamefont {Radice}}, \bibinfo
  {author} {\bibfnamefont {T.}~\bibnamefont {Radke}}, \bibinfo {author}
  {\bibfnamefont {C.}~\bibnamefont {Reisswig}}, \bibinfo {author}
  {\bibfnamefont {L.}~\bibnamefont {Rezzolla}}, \bibinfo {author}
  {\bibfnamefont {D.}~\bibnamefont {Rideout}}, \bibinfo {author} {\bibfnamefont
  {M.}~\bibnamefont {Ripeanu}}, \bibinfo {author} {\bibfnamefont
  {E.}~\bibnamefont {Schnetter}}, \bibinfo {author} {\bibfnamefont
  {B.}~\bibnamefont {Schutz}}, \bibinfo {author} {\bibfnamefont
  {E.}~\bibnamefont {Seidel}}, \bibinfo {author} {\bibfnamefont
  {E.}~\bibnamefont {Seidel}}, \bibinfo {author} {\bibfnamefont
  {J.}~\bibnamefont {Shalf}}, \bibinfo {author} {\bibfnamefont
  {U.}~\bibnamefont {Sperhake}}, \bibinfo {author} {\bibfnamefont
  {N.}~\bibnamefont {Stergioulas}}, \bibinfo {author} {\bibfnamefont {W.-M.}\
  \bibnamefont {Suen}}, \bibinfo {author} {\bibfnamefont {B.}~\bibnamefont
  {Szilagyi}}, \bibinfo {author} {\bibfnamefont {R.}~\bibnamefont {Takahashi}},
  \bibinfo {author} {\bibfnamefont {M.}~\bibnamefont {Thomas}}, \bibinfo
  {author} {\bibfnamefont {J.}~\bibnamefont {Thornburg}}, \bibinfo {author}
  {\bibfnamefont {M.}~\bibnamefont {Tobias}}, \bibinfo {author} {\bibfnamefont
  {A.}~\bibnamefont {Tonita}}, \bibinfo {author} {\bibfnamefont
  {P.}~\bibnamefont {Walker}}, \bibinfo {author} {\bibfnamefont {M.-B.}\
  \bibnamefont {Wan}}, \bibinfo {author} {\bibfnamefont {B.}~\bibnamefont
  {Wardell}}, \bibinfo {author} {\bibfnamefont {M.}~\bibnamefont {Zilhão}},
  \bibinfo {author} {\bibfnamefont {B.}~\bibnamefont {Zink}}, \ and\ \bibinfo
  {author} {\bibfnamefont {Y.}~\bibnamefont {Zlochower}},\ }\href {\doibase
  10.5281/zenodo.3522086} {\enquote {\bibinfo {title} {The {E}instein
  {T}oolkit},}\ } (\bibinfo {year} {2019}),\ \bibinfo {note} {to find out more,
  visit http://einsteintoolkit.org}\BibitemShut {NoStop}%
\bibitem [{\citenamefont {Schnetter}\ \emph {et~al.}(2004)\citenamefont
  {Schnetter}, \citenamefont {Hawley},\ and\ \citenamefont
  {Hawke}}]{Schnetter:2003rb}%
  \BibitemOpen
  \bibfield  {author} {\bibinfo {author} {\bibfnamefont {E.}~\bibnamefont
  {Schnetter}}, \bibinfo {author} {\bibfnamefont {S.~H.}\ \bibnamefont
  {Hawley}}, \ and\ \bibinfo {author} {\bibfnamefont {I.}~\bibnamefont
  {Hawke}},\ }\href {\doibase 10.1088/0264-9381/21/6/014} {\bibfield  {journal}
  {\bibinfo  {journal} {Class. Quant. Grav.}\ }\textbf {\bibinfo {volume}
  {21}},\ \bibinfo {pages} {1465} (\bibinfo {year} {2004})},\ \Eprint
  {http://arxiv.org/abs/gr-qc/0310042} {arXiv:gr-qc/0310042} \BibitemShut
  {NoStop}%
\bibitem [{\citenamefont {Thornburg}(2004)}]{Thornburg:2003sf}%
  \BibitemOpen
  \bibfield  {author} {\bibinfo {author} {\bibfnamefont {J.}~\bibnamefont
  {Thornburg}},\ }\href {\doibase 10.1088/0264-9381/21/2/026} {\bibfield
  {journal} {\bibinfo  {journal} {Class. Quant. Grav.}\ }\textbf {\bibinfo
  {volume} {21}},\ \bibinfo {pages} {743} (\bibinfo {year} {2004})},\ \Eprint
  {http://arxiv.org/abs/gr-qc/0306056} {arXiv:gr-qc/0306056} \BibitemShut
  {NoStop}%
\bibitem [{\citenamefont {Dreyer}\ \emph {et~al.}(2003)\citenamefont {Dreyer},
  \citenamefont {Krishnan}, \citenamefont {Shoemaker},\ and\ \citenamefont
  {Schnetter}}]{Dreyer:2002mx}%
  \BibitemOpen
  \bibfield  {author} {\bibinfo {author} {\bibfnamefont {O.}~\bibnamefont
  {Dreyer}}, \bibinfo {author} {\bibfnamefont {B.}~\bibnamefont {Krishnan}},
  \bibinfo {author} {\bibfnamefont {D.}~\bibnamefont {Shoemaker}}, \ and\
  \bibinfo {author} {\bibfnamefont {E.}~\bibnamefont {Schnetter}},\ }\href
  {\doibase 10.1103/PhysRevD.67.024018} {\bibfield  {journal} {\bibinfo
  {journal} {Phys. Rev. D}\ }\textbf {\bibinfo {volume} {67}},\ \bibinfo
  {pages} {024018} (\bibinfo {year} {2003})},\ \Eprint
  {http://arxiv.org/abs/arXiv:gr-qc/0206008} {arXiv:gr-qc/0206008} \BibitemShut
  {NoStop}%
\bibitem [{\citenamefont {Witek}\ \emph {et~al.}(2020)\citenamefont {Witek},
  \citenamefont {Zilhao}, \citenamefont {Ficarra},\ and\ \citenamefont
  {Elley}}]{Canuda_2020_3565475}%
  \BibitemOpen
  \bibfield  {author} {\bibinfo {author} {\bibfnamefont {H.}~\bibnamefont
  {Witek}}, \bibinfo {author} {\bibfnamefont {M.}~\bibnamefont {Zilhao}},
  \bibinfo {author} {\bibfnamefont {G.}~\bibnamefont {Ficarra}}, \ and\
  \bibinfo {author} {\bibfnamefont {M.}~\bibnamefont {Elley}},\ }\href
  {\doibase 10.5281/zenodo.3565475} {\enquote {\bibinfo {title} {{Canuda: a
  public numerical relativity library to probe fundamental physics}},}\ }
  (\bibinfo {year} {2020})\BibitemShut {NoStop}%
\bibitem [{EinsteinToolkit()}]{EinsteinToolkit:web}%
  \BibitemOpen
  EinsteinToolkit,\ \href@noop {} {\enquote {\bibinfo {title} {{Einstein
  Toolkit}: Open software for relativistic astrophysics},}\ }\bibinfo
  {howpublished} {\url{http://einsteintoolkit.org/}}\BibitemShut {NoStop}%
\bibitem [{Note1()}]{Note1}%
  \BibitemOpen
  \bibinfo {note} {Both the Einstein-Klein-Gordon and Einstein-Proca models
  introduce a single new parameter $\mu $ corresponding to the scalar or vector
  field mass. This scale is set to unity, so that masses, frequencies, etc, are
  given in units of $\mu $. Also geometrized units $G= 1 =c$ are
  used.}\BibitemShut {Stop}%
\bibitem [{\citenamefont {Hod}(2012)}]{Hod:2012px}%
  \BibitemOpen
  \bibfield  {author} {\bibinfo {author} {\bibfnamefont {S.}~\bibnamefont
  {Hod}},\ }\href {\doibase 10.1103/PhysRevD.86.129902,
  10.1103/PhysRevD.86.104026} {\bibfield  {journal} {\bibinfo  {journal}
  {Phys.Rev.}\ }\textbf {\bibinfo {volume} {D86}},\ \bibinfo {pages} {104026}
  (\bibinfo {year} {2012})},\ \Eprint {http://arxiv.org/abs/1211.3202}
  {arXiv:1211.3202 [gr-qc]} \BibitemShut {NoStop}%
\bibitem [{\citenamefont {Townsend}(1997)}]{Townsend:1997ku}%
  \BibitemOpen
  \bibfield  {author} {\bibinfo {author} {\bibfnamefont {P.}~\bibnamefont
  {Townsend}},\ }\href@noop {} {\  (\bibinfo {year} {1997})},\ \Eprint
  {http://arxiv.org/abs/gr-qc/9707012} {arXiv:gr-qc/9707012} \BibitemShut
  {NoStop}%
\bibitem [{Note2()}]{Note2}%
  \BibitemOpen
  \bibinfo {note} {See $e.g.$~\cite {sanchis2019head} for the definition of
  this potential.}\BibitemShut {Stop}%
\bibitem [{Note3()}]{Note3}%
  \BibitemOpen
  \bibinfo {note} {The larger $D$ guarantees a small constraints violation of
  the initial data.}\BibitemShut {Stop}%
\bibitem [{Note4()}]{Note4}%
  \BibitemOpen
  \bibinfo {note} {The initial mass/energy is larger than the mass of the BSs
  due to the boosts.}\BibitemShut {Stop}%
\bibitem [{Note5()}]{Note5}%
  \BibitemOpen
  \bibinfo {note} {As mentioned before spinning scalar boson stars are
  unstable~\cite {Sanchis-Gual:2019ljs}.}\BibitemShut {Stop}%
\end{thebibliography}%

\end{document}